\documentclass[aps,a4,amsmath,amssymb,floatfix]{revtex4-1}

\usepackage{bm}
\usepackage{graphicx}
\usepackage{amsbsy}
\usepackage{dcolumn}
\usepackage{slashbox}
\newcommand{\vlk}{$V_{\rm low-k}$ }
\newcommand{\vlkn}{$V_{\rm low-k}$}
\newcommand{\be}{\begin{equation}}
\newcommand{\ee}{\end{equation}}

\begin{document}

\title{Chiral Fermi liquid approach to neutron matter\footnote{Work 
supported in part by BMBF, the DFG cluster of excellence Origin and 
Structure of the Universe, and by DFG and NSFC (CRC110).}}

\author{J.\ W.\ Holt$^{1,3}$, N.\ Kaiser$^1$ and W.\ Weise$^{1,2}$}
\affiliation{$^1$Physik Department, Technische Universit\"{a}t M\"{u}nchen,
    D-85747 Garching, Germany}
\affiliation{$^2$ECT$^{\, *}$, Villa Tambosi, I-38123 Villazzano (TN), Italy}
\affiliation{$^3$Physics Department, University of Washington, Seattle,
    Washington 98195, USA}

\begin{abstract}

We present a microscopic calculation of the complete quasiparticle interaction,
including central as well as noncentral components, in neutron matter from 
high-precision two- and three-body forces derived within 
the framework of chiral effective field theory. The contributions 
from two-nucleon forces are computed in many-body perturbation theory
to first and second order (without any simplifying approximations). In 
addition we include the leading-order
one-loop diagrams from the N$^2$LO chiral three-nucleon force, which contribute
to all Fermi liquid parameters except those associated with the center-of-mass
tensor interaction. The relative-momentum dependence of the quasiparticle interaction 
is expanded in Legendre polynomials up to $L=2$. 
Second-order Pauli blocking and medium polarization effects 
act coherently in specific channels, namely for the Landau parameters $f_1$,
$h_0$ and $g_0$, which results in a dramatic increase in the quasiparticle effective
mass as well as a decrease in both the effective tensor force and the neutron 
matter spin susceptibility. For densities greater than about half nuclear matter
saturation density $\rho_0$, the contributions to the Fermi liquid parameters
from the leading-order chiral three-nucleon force scale in all cases 
approximately linearly with the nucleon density. The largest effect of the three-nucleon force is to generate  
a strongly repulsive effective interaction in the isotropic spin-independent channel.
We show that the leading-order 
chiral three-nucleon force leads to an increase in the spin susceptibility of neutron 
matter, but we observe no evidence for a ferromagnetic spin instability in the vicinity
of the saturation density $\rho_0$. This work sets the foundation for future 
studies of neutron matter response to weak and electromagnetic probes with 
applications to neutron star structure and evolution.

\end{abstract}

\maketitle
%\bigskip
%{\small PACS: 21.30.Fe, 21.60.Cs, 23.40.-s\\
%Keywords: Effective field theory at finite density, chiral three-nucleon
%force.}

%%%%%%%%%%%%%%%%%%%%%%%%%%%%%%%%%%%%%%%%%%%%%%%%%%%%%%%%%%%%%%%%%%%%%%%%%%%%%%%
%%%%%%%%%%%%%%%%%%%%%%%%%%%%%%%%%%%%%%%%%%%%%%%%%%%%%%%%%%%%%%%%%%%%%%%%%%%%%%%
\section{Introduction}

In a series of recent articles \cite{holt11,holt12}, we have revisited the 
Fermi liquid description of infinite nuclear matter in the context of modern
two- and three-nucleon interactions derived within the framework of
chiral effective field theory. In these studies we limited our discussion
to the central components of the quasiparticle interaction in a medium of
spin-saturated symmetric nuclear matter characterized by the nucleon density 
$\rho = 2k_f^3/3\pi^2$. In the vicinity of the saturation density
$\rho_0$, the central components of the quasiparticle
interaction are strongly constrained by the properties of bulk nuclear 
matter and its low-energy excitations. In refs.\ \cite{holt11,holt12} it was 
found that microscopic calculations of the quasiparticle interaction within 
many-body perturbation theory %with realistic nuclear forces 
yield an 
accurate description of the nuclear matter compressibility, isospin asymmetry 
energy and spin-isospin response only with the inclusion of leading-order 
medium effects, which arise at second order in a perturbative calculation with two-body 
forces and at first order in a calculation with three-nucleon forces.

In the present paper we extend these calculations to pure neutron matter, which
we treat as a normal (non-superfluid) Fermi liquid (the generalization of
Fermi liquid theory to 
superfluid Fermi systems has been performed in refs.\ 
\cite{legget66,leinson09,gusakov10}). Such a quasiparticle
description has been used in previous works to understand neutrino emissivity 
in neutron stars \cite{iwamoto82,navarro99,lykasov08,bacca09} as well as the spin
response of neutron star matter to strong magnetic fields \cite{haensel82,
olsson04,garcia09,pethick09}. In such calculations, the Landau parameters that 
characterize the quasiparticle interaction have been computed with microscopic
two-body nuclear forces or with phenomenological Skyrme and Gogny effective 
interactions. Qualitative differences arise between the predictions of 
microscopic and phenomenological forces, particularly with respect to 
the possibility of bulk magnetization or even the existence of a spontaneous 
ferromagnetic phase transition of neutron matter at several times nuclear matter
saturation density \cite{fantoni01,rios05,garcia09}. As discussed in Section
\ref{res}, three-neutron forces give rise to Fermi liquid parameters that scale
approximately linearly with the neutron density (at leading order), and
one of the primary goals of the present work is to better understand the role
of three-neutron correlations in describing the dynamical response of neutron 
matter to weak or electromagnetic probes.

Both the magnetic susceptibility of dense neutron matter as well as neutrino 
elastic scattering, absorption and emission rates are sensitive to the inclusion of 
noncentral components of the quasiparticle 
interaction \cite{olsson04,lykasov08,bacca09}, which have been introduced
in refs.\ \cite{haensel75,schwenk04}. Such interactions include (in the 
long-wavelength approximation) the exchange tensor interaction, proportional 
to $S_{12}(\hat q) = 3 \vec \sigma_1 \cdot \hat q\, \vec \sigma_2 \cdot \hat 
q -\vec \sigma_1 \cdot\vec \sigma_2$, which appears already in the free-space
nucleon-nucleon potential where it is dominated by one-pion exchange, as well
as center-of-mass tensor and cross-vector interactions, proportional to 
$S_{12}(\hat P)$ and $(\vec \sigma_1 \times \vec \sigma_2)\cdot 
(\hat q \times \hat P)$ respectively. In these spin-dependent operators, $\vec q = 
\vec p_1 - \vec p_2$ is the relative momentum and $\vec P = \vec p_1 + \vec p_2$
the center-of-mass momentum of the two quasiparticles
on the Fermi surface $(|\vec p_{1,2}| = k_f)$. When 
including effects of the medium in the form of loop integrals over the filled Fermi 
sea of neutrons, all noncentral interactions can be generated. In fact, the 
second-order contributions from two-body forces produce exchange tensor,
center-of-mass tensor and cross-vector interactions. 
Note that the spin-nonconserving cross-vector interaction can 
only be generated through polarization (particle-hole) corrections to the 
effective interaction and not by ladders \cite{schwenk04}. To date there has been no study 
of three-nucleon force contributions to either the center-of-mass tensor
or cross-vector interactions, while in ref.\ \cite{kaiser06} loop diagrams 
involving intermediate $\Delta$-isobar excitations were analyzed and shown 
to generate an exchange tensor interaction in symmetric nuclear matter.  

The present work sets the foundation for a systematic study of the role 
played by second-order corrections and three-nucleon forces in determining the noncentral components of 
the quasiparticle interaction. Their effect on neutron star structure
and evolution will be studied in future work. We employ many-body 
perturbation theory and compute all contributions to the quasiparticle 
interaction for the case of two neutrons on the Fermi surface interacting in a background medium
of pure neutron matter. We employ high-precision chiral two- and three-nucleon
interactions whose unresolved short-distance components are encoded in a set of
contact couplings proportional to low-energy constants fit to elastic
nucleon-nucleon scattering phase shifts and properties of light nuclei
\cite{entem03,epelbaum06,gazit09}. To improve the convergence of the microscopic
calculation of the quasiparticle interaction in many-body perturbation 
theory, we employ renormalization-group methods for integrating out
momenta beyond the scale of $\Lambda \simeq 2.0$\,fm$^{-1}$, which results
in the nearly universal two-nucleon potential \vlk \cite{bogner03,bogner10}. 
The contributions to the quasiparticle interaction from the leading-order 
chiral three-nucleon force (with scale-dependent low-energy constants) is 
computed to one-loop order.

The present paper is organized as follows. In Section \ref{qpnm} we give a brief review
of Landau's theory of normal Fermi liquids and discuss the microscopic
approach for computing the quasiparticle interaction. We then present a general
method that can be applied to any free-space nucleon-nucleon force
given in a partial-wave basis for extracting the scalar functions that multiply 
the various spin-dependent operators (namely the central, spin-spin, exchange 
tensor, center-of-mass tensor and cross-vector terms) occuring in the quasiparticle 
interaction. We then discuss the microscopic origin of the spin-nonconserving 
cross-vector interaction, which at second-order results exclusively from the interference of a 
two-body spin-orbit force with (in principle) any non-spin-orbit 
component in the free-space nucleon-nucleon interaction. In Section \ref{mi} we 
test the intricate (numerical) calculations of the second-order contributions to the 
quasiparticle interaction by means of model interactions that can be solved partially 
analytically. In Section \ref{lp3nf} we present analytical formulas for the Landau parameters
of the quasiparticle interaction arising from the leading-order N$^2$LO chiral 
three-neutron force. The results of the corresponding calculations with two- and three-neutron
forces at several resolution scales are presented and 
discussed in Section \ref{res}. We end with a summary and conclusions.

%\begin {figure}
%\begin{center}
%\end{center}
%\end {figure}
%\input{blub.tex}

%%%%%%%%%%%%%%%%%%%%%%%%%%%%%%%%%%%%%%%%%%%%%%%%%%%%%%%%%%%%%%%%%%%%%%%%%%%%%%%
%%%%%%%%%%%%%%%%%%%%%%%%%%%%%%%%%%%%%%%%%%%%%%%%%%%%%%%%%%%%%%%%%%%%%%%%%%%%%%%
\section{Quasiparticle interaction in neutron matter}
\label{qpnm}
\subsection{General structure of the quasiparticle interaction and spin-space decomposition}
\label{gs}

In Landau's theory of normal Fermi liquids \cite{landau57,baym91}, the quasiparticle interaction
${\cal F}(\vec p_1 s_1 t_1; \vec p_2 s_2 t_2)$ 
derives from the change in the total energy density due to second-order variations
in the particle occupation densities $\delta n_{\vec p, s, t}$:
\be
\delta {\cal E} = \sum_{\vec p s t} \epsilon_{\vec p}\, \delta 
n_{\vec p s t} + \frac{1}{2} \sum_{\substack{{\vec p}_1 s_1 t_1
 \\ {\vec p}_2 s_2 t_2}}{\cal F}({\vec p}_1 s_1 t_1;
{\vec p}_2 s_2 t_2) \delta n_{{\vec p}_1 s_1 t_1}
\delta n_{{\vec p}_2 s_2 t_2} + \cdots,
\label{deltae}
\ee
where $s_i$ and $t_i$ label the spin and isospin quantum numbers of 
the $i$th quasiparticle.
The most general form for the effective interaction between two quasiparticles
in pure neutron matter in the long-wavelength limit is given by \cite{schwenk04}
\begin{eqnarray}
{\cal F}(\vec p_1, \vec p_2\,) &=& f(\vec p_1, \vec p_2\,) + g(\vec p_1, 
\vec p_2\,) \vec \sigma_1 \cdot \vec \sigma_2 + h (\vec p_1, \vec p_2\,) 
S_{12}(\hat q) + k (\vec p_1, \vec p_2\,) S_{12}(\hat P) \nonumber \\ 
&& +l (\vec p_1, \vec p_2\,) (\vec \sigma_1 \times \vec \sigma_2)\cdot 
(\hat q \times \hat P),
\label{qpi}
\end{eqnarray}
where $\vec q = \vec p_1 - \vec p_2$ is the momentum transfer in the 
exchange channel, $\vec P = \vec p_1 + \vec p_2$ is the conserved
center-of-mass momentum and the tensor operator $S_{12}(\hat v)$ is defined by
$S_{12}(\hat v) = 3 \vec \sigma_1 \cdot \hat v\, \vec
\sigma_2 \cdot \hat v -\vec \sigma_1 \cdot\vec \sigma_2$. 
The interaction in eq.\ (\ref{qpi}) is invariant
under parity and time-reversal transformations as well as under
the interchange of the particle labels. However, due to the presence of the
medium, Galilean invariance is no longer manifest, leading to new operator
structures (namely $S_{12}(\hat P)$ and $A_{12}(\hat q, \hat P) = 
(\vec \sigma_1 \times \vec \sigma_2)\cdot 
(\hat q \times \hat P)$) that depend explicitly on the center-of-mass momentum $\vec P$. 
Since the two quasiparticle momenta lie on the Fermi surface $|\vec p_1\,| = |\vec p_2\,| = k_f$, 
the remaining angular dependence of the
quasiparticle interaction is conveniently expanded in Legendre polynomials of
$\cos \theta ={\hat p}_1 \cdot {\hat p}_2$:
\begin{equation}
\chi({\vec p}_1,{\vec p}_2) = \sum_{L=0}^\infty \chi_L(k_f) P_L(\mbox{cos } \theta),
\label{gflp}
\end{equation}
where $\chi$ represents $f, g, h, k,$ or $l$, and the angle $\theta$ is
related to the relative momentum $q = |{\vec p}_1 - {\vec p}_2|$ 
through the relation $q = 2k_f\, {\rm sin}\, (\theta /2)$.

The Fermi liquid parameters of nuclear matter can be either extracted from experiment
(due to direct relations between particular Landau parameters and nuclear observables 
\cite{migdal67,baym91}) or computed microscopically within many-body perturbation 
theory \cite{abrikosov59}. Given a strongly interacting normal Fermi system at low
temperatures, it is not clear that a pertubative approach is justified. However,
in the case of nuclear and neutron matter, there are strong indications that renormalization group
techniques \cite{bogner03, bogner10, bogner02} may render nuclear interactions perturbative 
for a wide range of densities when evolved down to a cutoff scale $\Lambda \alt 
2.1$\,fm$^{-1}$. In the framework of many-body perturbation theory, the quasiparticle 
interaction is extracted by functionally differentiating the contributions to the 
ground-state energy density twice with respect to the particle occupation numbers. Previous 
work \cite{holt11,holt12} has
shown that a satisfactory description of bulk nuclear matter properties around the
saturation density can be obtained with chiral and low-momentum nuclear 
interactions only with the inclusion of the leading-order Pauli-blocking and 
medium-polarization effects from two- and three-nucleon interactions. This 
requires a second-order perturbative calculation in the case of two-nucleon
interactions together with the first-order perturbative contribution from
three-nucleon forces. These systematic calculations suggest that a consistent microscopic description
of the quasiparticle interaction in neutron matter, for which there is much less 
empirical information, can be achieved.

The relations between specific Fermi liquid parameters and bulk properties of 
the medium are well known \cite{migdal67,baym91}. The compression modulus of neutron
matter (at density $\rho = k_f^3/3\pi^2$)
\be
{\cal K}=\frac{3k_F^2}{M^*} \left (1+F_0\right ),
\label{comp}
\ee
is related to the Fermi liquid parameter $F_0 = N_0 f_0$ (with $N_0 = k_f M^*/\pi^2$ the density of states
at the Fermi surface) which represents the isotropic 
part of the spin-independent quasiparticle interaction. 
Note that the compressibility ${\cal K} = k_f^2 \, \partial^2 \bar E/\partial k_f^2 + 4k_f\, \partial \bar
E/\partial k_f$ is determined by both the curvature and slope of the energy per particle $\bar E$.
In eq.\ (\ref{comp}) the 
quasiparticle effective mass $M^*$ is given by
\be
\frac{M^*}{M_n} = 1+\frac{F_1}{3},
\label{effmass}
\ee
with $M_n=939.6$\,MeV the free neutron mass and $F_1 = N_0 f_1$. 
Considering only the central components of the quasiparticle
interaction, the neutron matter spin susceptibility is given by 
\be
\chi = \mu_n^2 \frac{N_0}{1+G_0},
\label{susc}
\ee
where $\mu_n=-1.913$ is the free-space neutron magnetic moment (in units of the nuclear magneton)
and $G_0 = N_0 g_0$. The presence of noncentral 
components in the quasiparticle interaction that couple quasiparticle spins
to their momenta results in effective charges (magnetic moments) that are not
scalars under rotations of the quasiparticle momentum. The resulting expression
for the spin susceptibility then involves both longitudinal and transverse
components of the magnetic moment \cite{haensel82,olsson04}. 

The first calculation to extract all components of the quasiparticle interaction
given in eq.\ (\ref{qpi}) for an arbitrary two-neutron force was performed in ref.\
\cite{schwenk04}. In the following we present a general method to project out 
the various momentum-dependent scalar functions $f,g,h,k$ and $l$ of the 
quasiparticle interaction in eq.\ (\ref{qpi}). This is achieved by taking 
specific linear combinations of the spin-space matrix
elements of the quasiparticle interaction. The form of this matrix will of course depend on the 
choice of coordinate system. For $\vec q =
\vec p_1 - \vec p_2 = q\, \vec e_z$ and $\vec P = \vec p_1 + \vec p_2 = P \vec e_x$,
the spin-space matrix elements $\langle m_s | {\cal F}(\vec p_1, \vec p_2) | m_s^\prime \rangle$ 
in terms of the scalar functions $f,g,h,k,$ and $l$ are given by
\begin{equation}
\begin{array}{c||ccc|c}
\multicolumn{1}{c||}{m_s \backslash m_s^\prime} & \multicolumn{1}{c}{1} & 
\multicolumn{1}{c}{0} & \multicolumn{1}{c}{-1} & \multicolumn{1}{c}{0} \\ \hline \hline %\cline{2-5}
1 & f+g+2h-k & 0 & 3k & \sqrt{2} l \\
0 & 0 & f+g-4h+2k & 0 & 0 \\
-1 & 3k & 0 & f+g+2h-k & \sqrt{2} l \\ \cline{2-5}
0 & \sqrt{2} l & 0 & \sqrt{2} l & f-3g
\end{array}.
\label{zproj}
\nonumber
\end{equation}
The upper left $3\times 3$ submatrix gives the nine triplet-triplet matrix elements,
while the fourth row and fourth column give the matrix elements that include the singlet state. 
In this coordinate frame, the exchange tensor interaction $h(\vec p_1, \vec p_2\,) 
S_{12}(\hat q)$ gives nonzero contributions
only for the diagonal triplet spin-space matrix elements. The center-of-mass tensor force
$k(\vec p_1, \vec p_2\,)S_{12}(\hat P)$ contributes to the triplet diagonal matrix 
elements as well as the matrix elements
mixing $m_s = \pm 1$ with $m_s^\prime = \mp 1$. The cross-vector
interaction $l(\vec p_1, \vec p_2\,)A_{12}(\hat q, \hat P)$ is nonvanishing only in spin-nonconserving transitions with
 $|m_s-m_s^\prime| = 1$. We note that it is not possible to separate the exchange tensor from the 
center-of-mass tensor interaction when considering only diagonal spin-space
matrix elements, since they appear in the same linear combination $2h-k$.
An alternative choice of coordinate system in which $\vec q=q\, \vec e_x$ and $\vec P = P \vec e_z$ 
just interchanges the contributions of the exchange tensor and center-of-mass
tensor operators (and $l$ comes with opposite sign).

From the above matrix, the scalar functions multiplying the spin-dependent 
operators are extracted as the following linear combinations of the 
spin-space matrix elements:
\begin{eqnarray}
f&=&(2{\cal F}_{1,1}^t+{\cal F}_{0,0}^t+{\cal F}_{0,0}^s)/4\,, \nonumber \\
g&=&(2{\cal F}_{1,1}^t+{\cal F}_{0,0}^t-3{\cal F}_{0,0}^s)/12\,, \nonumber \\
h&=&({\cal F}_{1,1}^t+{\cal F}_{1,-1}^t-{\cal F}_{0,0}^t)/6\,, \nonumber \\
k&=&({\cal F}_{1,-1}^t)/3\,, \nonumber \\
l&=&({\cal F}_{1,0}^{ts})/\sqrt{2}\,,
\label{projform}
\end{eqnarray}
where the superscripts $s,t$ distinguish spin-singlet and spin-triplet states
of two quasiparticles, and the two subscripts label $m_s$
and $m_s^\prime$, respectively. Only the function $l(\vec p_1,
\vec p_2\,)$ multiplying the cross-vector operator $A_{12}(\hat q, \hat P)$ depends on a singlet-triplet 
mixing matrix element.

\subsection{First-order contribution}

With the relations given in eq.\ (\ref{projform}) we need to compute only the 
quasiparticle interaction in different total spin states. 
The first-order perturbative contribution to the energy density is given by
\be
{\cal E}^{(1)}_{2N} = \frac{1}{2}\sum_{12} \langle {\vec k}_1 s_1; {\vec k}_2 s_2
| \bar V | {\vec k}_1 s_1; {\vec k}_2 s_2  \rangle n_1 n_2,
\ee
where $\bar V = (1-P_{12})V$ denotes the antisymmetrized two-body potential,
$n_j=\theta(k_f-|\vec k_j|)$ is the usual zero-temperature Fermi distribution and 
the summation includes both spin and momenta. The corresponding contribution
to the quasiparticle interaction reads
\begin{equation}
{\cal F}^{(1)}_{2N}({\vec p}_1 s_1; {\vec p}_2 s_2)
 = \langle {\vec p}_1 s_1; {\vec p}_2 s_2 | 
\bar V | {\vec p}_1 s_1 ; {\vec p}_2 s_2 \rangle
\equiv \langle 12 | \bar V | 12 \rangle.
\label{order1qp}
\end{equation}
Setting the relative momentum $\vec q = \vec p_1 -\vec p_2$ 
along the $\vec e_z$-direction and projecting onto Legendre polynomials $P_L(\hat p_1
\cdot \hat p_2)$, the 
Fermi liquid parameters are obtained from eq.\ (\ref{order1qp}) in terms of the matrix
elements of $V$ depending on $p=q/2$ in a partial-wave representation:
\begin{eqnarray}
{\cal F}_L (k_f; S m_s m_s^\prime)&=& 2(2L+1) \sum_{l l^\prime J}
i^{l - l^\prime} (1+(-1)^{l+S}) \sqrt{(2l+1)(2 l^\prime +1)} 
\, \langle l 0 \, S m_s | J M \rangle \nonumber \\ 
&&\times  \langle l^\prime 0 \, S m_s^\prime
| JM \rangle \int_0^{k_f} dp\, \frac{p}{k_f^2}  \, \langle plSJM | V | 
pl^\prime SJM \rangle P_L(1-2p^2/k_f^2),
\label{qp1st}
\end{eqnarray}
where we are following the normalization convention in ref.\ \cite{holt11}.
The leading-order contribution is simply a kinematically-restricted form of the free-space interaction, which
contains no center-of-mass dependent components. Therefore, the only nonzero
contributions to the quasiparticle interaction are $f,g,$ and $h$. This is further
reflected in the partial wave decomposition through the obvious property
$M=m_s=m_s^\prime$ implied by the Clebsch-Gordan coefficients in eq.\ (\ref{qp1st}).

\begin{figure}
\begin{center}
\includegraphics[height=4cm]{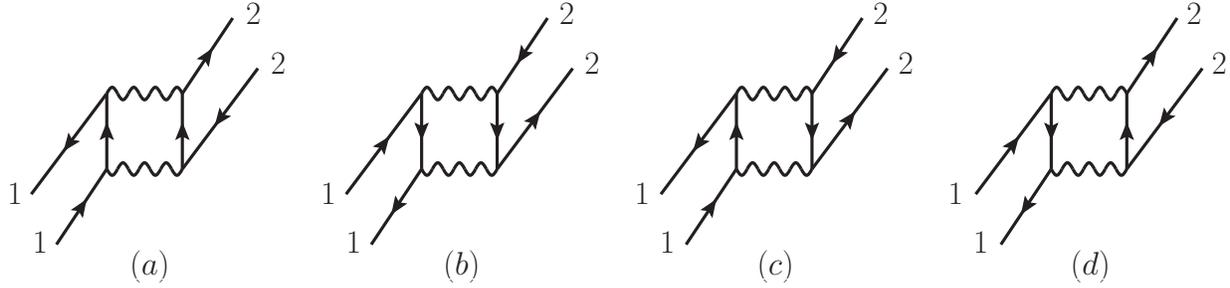}
\end{center}
\vspace{-.5cm}
\caption{Diagrams contributing to the second-order quasiparticle interaction (all 
interactions represented by wavy lines are antisymmetrized): (a) particle-particle diagram, 
(b) hole-hole diagram, and (c)+(d) particle-hole diagrams.}
\label{pphhph}
\end{figure}

\subsection{Second-order contributions}
\label{soc}

The contribution to the energy density from the two-neutron force at 
second-order in many-body perturbation theory has the form
\be
{\cal E}^{(2)}_{2N} = \frac{1}{4}\sum_{1234}
\frac{|\langle 12 | \bar V | 34 \rangle|^2\, n_1 n_2 (1-n_3)
(1-n_4)} {\epsilon_1 + \epsilon_2 - \epsilon_3 - \epsilon_4},
\ee
where in a plane-wave basis $\epsilon_j=\vec k_j^2/2M_n^{(*)}$ is the 
single-particle energy associated with the momentum $\vec k_j$.
Functionally differentiating twice with respect to the quasiparticle 
distribution functions yields four different contributions to the
quasiparticle interaction:
\begin{eqnarray}
{\cal F}^{(2)}_{2N} ({\vec p}_1 s_1 t_1; {\vec p}_2 s_2 t_2)
&=& \frac{1}{2} \sum_{34} \frac{|\langle 12 | \bar V | 34 \rangle|^2 (1-n_3)
(1-n_4)} {\epsilon_1 + \epsilon_2 - \epsilon_3 - \epsilon_4} 
+\frac{1}{2} \sum_{34} \frac{|\langle 12 | \bar V | 34 \rangle |^2 n_3 n_4}
{\epsilon_3 + \epsilon_4 - \epsilon_1 - \epsilon_2}  \nonumber \\
 &-& \sum_{34} \frac{|\langle 13 | \bar V | 24 \rangle |^2 n_3 (1-n_4)}
{\epsilon_1 + \epsilon_3 - \epsilon_2 - \epsilon_4} - \sum_{34} 
\frac{|\langle 13 | \bar V | 24 \rangle |^2 n_4 (1-n_3)}
{\epsilon_1 + \epsilon_4 - \epsilon_2 - \epsilon_3}.
\label{order2qp}
\end{eqnarray}
A graphical representation of these terms is given in
Fig.\ \ref{pphhph}. The first two terms in eq.\ (\ref{order2qp}) are called
the particle-particle ($pp$) and hole-hole ($hh$) contributions, which have a very similar 
structure. The particle-particle contribution is given by 
\begin{eqnarray}
{\cal F}^{pp}(\vec p_1 \vec p_2;S m_s S^\prime m_s^\prime)
&=& \frac{1}{2} \sum_{\bar S \bar m_s} \int \frac{d^3k_3}{(2\pi)^3} 
\frac{d^3k_4}{(2\pi)^3} \frac{ \langle \vec p_1 \vec p_2 S m_s 
| \bar V | \vec{k}_3 \vec{k}_4 \bar S \bar m_s \rangle \langle 
\vec k_3 \vec k_4 \bar S \bar m_s | \bar V | \vec p_1 \vec p_2 
S^\prime m_s^\prime \rangle} {\epsilon_{p_1} + \epsilon_{p_2} -
\epsilon_{k_3} - \epsilon_{k_4}
} \nonumber \\
&\times & (1-n_3)(1-n_4) (2\pi)^3 \delta(\vec p_1+ 
\vec p_2-\vec{k}_3- \vec{k}_4),
\label{ppdia}
\end{eqnarray}
where we allow for the possibility that $m_s\ne m_s^\prime$. The hole-hole contribution is 
easily obtained from eq.\ (\ref{ppdia}) by changing the sign of the 
energy denominator and replacing the two particle distribution functions
$(1-n_3)(1-n_4)$ with hole distribution functions $n_3n_4$.
In computing the $pp$ and $hh$ diagrams, we found it more 
convenient to align the total momentum $\vec P$ along the $\vec e_z$-direction and 
the relative momentum $\vec q$ along the $\vec e_x$-direction. For this choice of
coordinate system $h$ and
$k$ are just interchanged in eq.\ (\ref{zproj}). In a partial-wave basis the particle-particle
contribution in eq.\ (\ref{ppdia}) reads
\begin{eqnarray}
&&{\cal F}^{pp}_L(S m_s m_s^\prime) = \frac{2L+1}{4 \pi^2 k_F^2} \sum_{\substack {l_1l_2l_3l_4mm^\prime \\ 
\bar m \bar m_s JJ^\prime M}} \int_0^{k_F} dp \, p \int_p^{\infty} dk \, k^2 N(l_1ml_2\bar m l_3m^\prime l_4 \bar m) 
P_{l_1}^m(0) P_{l_3}^{m^\prime}(0) \nonumber \\
&& \hspace{.3in} \times   \frac{M_n}{p^2-k^2} i^{l_2+l_3-l_1-l_4} {\cal C}^{JM}_{l_1m S m_s} 
{\cal C}^{JM}_{l_2\bar m S \bar m_s} 
{\cal C}^{J^\prime M}_{l_3 m^\prime S m_s^\prime} {\cal C}^{J^\prime M}_{l_4 \bar m S \bar m_s}  
\int^{{\rm min}\{ x_0,1 \}}_{{\rm max}\{ -x_0,-1 \}} 
d x \, P_{l_2}^{\bar m}(x) P_{l_4}^{\bar m}(x) \nonumber \\
&& \hspace{.3in} \times  \langle pl_1SJM| \bar V | kl_2SJM \rangle
\langle kl_4SJ^\prime M | \bar V | pl_3SJ^\prime M \rangle P_L(1-2p^2/k_F^2),
\label{pp2nd}
\end{eqnarray}
where $P_l^m$ are the associated Legendre functions, $x=\cos \theta_k$, $\vec p = (\vec p_1 -\vec p_2)/2$, 
$\vec k = (\vec k_3 - \vec k_4)/2$, 
$x_0 = (k^2-p^2)/(2k\sqrt{k_F^2-p^2})$ and $N(l_1 m l_2\bar m l_3 m^\prime l_4 \bar m) = N_{l_1}^m
N_{l_2}^{\bar m} N_{l_3}^{m^\prime} N_{l_4}^{\bar m}$ with $N_l^m = \sqrt{(2l+1)(l-m)!/(l+m)!}$.
For $m_s=m_s^\prime$, this expression agrees with eq.\ (21) in ref.\ \cite{holt11}. 
From the underlying parity invariance of the two-neutron interaction, 
which preserves the total spin $S$, the $pp$ and $hh$ contributions cannot
give rise to the spin-nonconserving cross-vector interaction. However, spin-orbit and tensor
terms in the free-space neutron-neutron interaction do not conserve $m_s$ and
therefore can, at second-order in the $pp$ and $hh$ diagrams, give rise to a 
center-of-mass tensor interaction. This mechanism is exemplified in Section \ref{mi}
with spin-orbit and tensor two-body model interactions.

We split the second-order particle-hole ($ph$) contribution into two
pieces ${\cal F}^{(c)}_{ph} + {\cal F}^{(d)}_{ph}$ (see Fig.\ \ref{pphhph}). 
The first term arises from a coupling of the incoming or outgoing quasiparticle 1 to a hole state, while 
for the second term quasiparticle 1 couples to a particle state. The explicit expression reads
\begin{eqnarray}
{\cal F}_{ph}^{(c)}(\vec p_1\vec p_2;s_1 s_2 s_1^\prime s_2^\prime)
&=& \sum_{s_3 s_4} \int \frac{d^3k_3}{(2\pi)^3} \frac{d^3k_4}{(2\pi)^3}
\frac{ \langle \vec p_1 \vec k_3 s_1 s_3 | \bar V | \vec p_2 \vec k_4
 s_2 s_4 \rangle \langle \vec p_2 \vec k_4 s_2^\prime s_4 | 
\bar V | \vec p_1 \vec k_3 s_1^\prime s_3 \rangle}
{\epsilon_{p_2}+ \epsilon_{k_4}- \epsilon_{p_1} - \epsilon_{k_3}} \nonumber \\
&\times& n_3(1-n_4)
(2\pi)^3 \delta(\vec p_1+\vec k_3 - \vec p_2- \vec k_4),
\label{phdiaa}
\end{eqnarray}
where we allow for the possibility that the spin states of 
an incoming and outgoing quasiparticle can be different. 
In contrast to the treatment of the $ph$ contribution to the central components
 of the quasiparticle interaction described in ref.\ \cite{holt11}, 
here we have found it convenient to set $\vec p_1 - \vec p_2 
= q\, \vec e_z$ and $\vec p_1 + \vec p_2 = P \vec e_x$. In this case we write
\begin{eqnarray}
{\cal F}_L^{ph(c)}(s_1 s_2 s_1^\prime s_2^\prime) &=& \frac{2L+1}{32\pi^3} \int_{-1}^{1} d \cos \theta
P_L(\cos \theta) \int_0^{2\pi} d \phi_3 \int_{{\rm max}\{0, y_0\}}^{k_f}
dk_3  k_3^2  \int_{{\rm max}\{-1,z_0 \}}^1 d \cos \theta_3  
 \nonumber \\ 
&& \hspace{-.9in} \times \sum_{\substack {l_1l_2l_3l_4s_3 s_4 \\ 
m_1 m_2 m_3 m_4}} 
\langle p^\prime l_1m_1s_1 s_3| \bar V | k^\prime l_2 m_2 s_2 s_4 \rangle 
\langle k^\prime l_4 m_4 s_2^\prime s_4 | \bar V | p^\prime l_3 m_3 s_1^\prime s_3 \rangle \nonumber \\
&& \hspace{-.9in} \times \cos ((m_3 - m_1 + m_2 - m_4)\phi_{p^\prime})
P_{l_1}^{m_1}(\cos \theta_{p^\prime}) P_{l_2}^{m_2}(\cos \theta_{k^\prime})
P_{l_3}^{m_3}(\cos \theta_{p^\prime}) P_{l_4}^{m_4}(\cos \theta_{k^\prime}) \nonumber \\
&& \hspace{-.9in} \times i^{l_2+l_3-l_1-l_4}  N(l_1 m_1 l_2 m_2 l_3 m_3 l_4 m_4) 
\frac{M_n}{k_fk_3\cos \theta_3 \sin \theta/2 +k_f^2\sin^2\theta/2},
\label{pha}
\end{eqnarray}
where $\vec p^{\, \prime} = (\vec p_1 - \vec k_3)/2$, $\vec k^\prime = (\vec p_2 - \vec k_4)/2$,
$y_0 = k_f(1-2\sin \theta/2)$ and
$z_0 = (k_f^2-k_3^2-4k_f^2 \sin^2\theta/2)/(4k_fk_3\sin\theta/2)$. 
The product of exponentials coming from the spherical harmonics has been 
simplified to a cosine by noting that the imaginary part vanishes and that 
$\phi_{p^\prime} = \phi_{k^\prime}$. The expression
in eq.\ (\ref{pha}) can be further written out in terms of partial-wave matrix elements
by first coupling to total spin $S$ and then total angular momentum $J$.
The extraction of the scalar functions $f,g,h,k$ and $l$ is achieved
by taking appropriate linear combinations of the sixteen spin-space matrix elements
${\cal F}_L(s_1 s_2 s_1^\prime s_2^\prime)$, namely,
\be
\langle S m_s | {\cal F}_L^{ph} | S^\prime m_s^\prime \rangle = 
\sum_{s_1 s_1^\prime s_2 s_2^\prime} {\cal C}_{\frac{1}{2}s_1 \frac{1}{2} s_2^\prime}^{S m_s}
{\cal C}_{\frac{1}{2} s_1^\prime \frac{1}{2} s_2}^{S^\prime m_s^\prime}
{\cal F}_L^{ph}(s_1 s_2 s_1^\prime s_2^\prime).
\label{ssph}
\ee
It is a good check of the calculation that the resulting spin-space matrix on the left-hand
side of eq.\ (\ref{ssph}) is of the form introduced in subsection \ref{gs}.

The particle-hole polarization contribution can give rise to all 
noncentral interactions, as pointed out first in ref.\ \cite{schwenk04}.
However, the microscopic origin of the spin-nonconserving cross-vector interation should
be clarified. In fact, neither tensor forces nor spin-orbit
forces alone, when iterated in the particle-hole channel, generate
the cross-vector interaction. This can be seen from the spin structure of
eq.\ (\ref{phdiaa}). In order to produce a nonvanishing singlet-triplet mixing
matrix element, we can consider without loss of generality the spin-flip transition 
\begin{equation}
\langle \uparrow \downarrow | {\cal F}_{ph}^{(c)}(\vec p_1\vec p_2) | 
\uparrow \uparrow \rangle \sim 
\sum_{s_3 s_4} \langle \vec p_1 \vec k_3 \uparrow s_3 |
\bar V | \vec p_2 \vec k_4 \uparrow s_4 \rangle \langle \vec p_2 \vec k_4
\downarrow s_4 | \bar V | \vec p_1 \vec k_3 \uparrow s_3 \rangle.
\label{cvss}
\end{equation}
It is easily shown with momentum conservation ($\vec p_1 +\vec k_3 = \vec p_2 + \vec k_4$)
 that $\vec p^{\, \prime} - \vec k^\prime = \vec p_1 - \vec p_2 = q\, \vec e_z$.
In this case, the free-space tensor force ($\sim 3 \sigma_1^z \sigma_2^z - \vec \sigma_1 \cdot
\vec \sigma_2$) vanishes for $|\Delta m_s| = 1$, 
and therefore one of the two matrix elements in eq.\ (\ref{cvss}) will vanish for any values of 
$s_3$ and $s_4$. Similarly, for a free-space spin-orbit interaction of
the form $ i V_{so}(\vec \sigma_1 + \vec \sigma_2) \cdot (\vec p^{\, \prime} \times
\vec k^\prime \,)$, the vector $\vec p^{\, \prime} \times \vec k^\prime$ lies in the $x-y$ plane
and in this case the spin-orbit operator $(\vec \sigma_1 + \vec \sigma_2) \cdot (\vec p^{\, \prime} \times
\vec k^\prime \,)$ has very restricted matrix elements which are nonvanishing only in 
spin-triplet states with $|\Delta m_s|
= 1$. Again, one of the matrix elements in eq.\ (\ref{cvss}) will vanish for
any possible values of $s_3$ and $s_4$. These arguments indicate that at second-order
only the interference of a spin-orbit interaction with any other (non-spin-orbit) component can produce
the cross-vector interaction. In the following section, we will demonstrate that in fact 
central, spin-spin and tensor components all give nonvanishing interference terms.
Finally, we point out that although individually
${\cal F}^{(c)}_{ph}$ and ${\cal F}^{(d)}_{ph}$ can have forbidden
matrix elements for $|\Delta m_s| = 1$ in the triplet states, when summed 
together such terms cancel exactly. All other spin-space matrix elements of 
${\cal F}^{(c)}_{ph}$ and ${\cal F}^{(d)}_{ph}$ are
identical. These features serve as a good check of the 
involved calculation of the quasiparticle interaction at second order.

\subsection{Contribution from chiral three-neutron forces}
\label{ctnf}

Finally we consider the leading-order contribution from the N$^2$LO chiral three-nucleon
 force \cite{epelbaum06} in pure neutron matter. Only the components of the 
two-pion exchange three-nucleon force proportional to the low-energy constants 
$c_1$ and $c_3$ remain for neutrons:
\begin{equation}
V_{3n}^{(2\pi)} = \sum_{i\neq j\neq l} \frac{g_A^2}{4f_\pi^4} 
\frac{\vec{\sigma}_i \cdot \vec{q}_i \, \vec{\sigma}_j \cdot
\vec{q}_j}{(\vec{q_i}^2 + m_\pi^2)(\vec{q_j}^2+m_\pi^2)}
\left (-2c_1m_\pi^2
 + c_3 \vec{q}_i \cdot \vec{q}_j \right ) ,
\label{3n1}
\end{equation}
with parameters $g_A=1.29$, $f_\pi = 92.4$\,MeV and $m_{\pi} = 138$\,MeV (average
pion mass). The quantity $\vec{q}_i$ is the difference between the final and initial 
momenta of neutron $i$. In the following we employ two different choices for the 
low energy constants $c_1$ and $c_3$ in eq.\ (\ref{3n1}). When combined with the 
bare chiral N$^3$LO nucleon-nucleon potential \cite{entem03} we choose the values
$c_1 =-0.81\,$GeV$^{-1}$ and $c_3=-3.2\,$GeV$^{-1}$, whereas with the low-momentum
nucleon-nucleon potential \vlk we use $c_1 =-0.76\,$GeV$^{-1}$ and $c_3=-4.78\,$GeV$^{-1}$
\cite{nogga04,rentmeester03}. This variation in the low-energy constants (as well as the resolution scale $\Lambda$) 
provides a means for assessing theoretical uncertainty at a given order in many-body perturbation
theory.

The first-order contribution to the energy density of
neutron matter has the form
\be
{\cal E}^{(1)}_{3n} = \frac{1}{6}{\rm Tr}_{\sigma_i,\sigma_j,\sigma_l}
\int \frac{d^3k_i}{(2\pi)^3} \frac{d^3k_j}{(2\pi)^3} \frac{d^3k_l}{(2\pi)^3}
n_i n_j n_l \langle ijl | \bar{V}_{3N} | ijl 
\rangle \, ,
\ee
where $\bar{V}_{3n} = V_{3n}(1-P_{12}-P_{23}-P_{13}+P_{12}P_{23}+P_{13}P_{23})$
is the fully antisymmetrized three-neutron interaction and $n_j
=\theta(k_f-|\vec k_j|)+(2\pi)^3 \delta^3(\vec k_j-\vec p_j)\, 
\delta n_{\vec p_j \sigma_j}$. 
Functionally differentiating twice with respect to the two quasiparticle 
distribution functions yields
\be
{\cal F}^{(1)}_{3n}({\vec p}_1, {\vec p}_2) = \frac{1}{2}{\rm Tr}_{\sigma_i}
\int \frac{d^3k_i}{(2\pi)^3} n_i \langle i12 | \bar{V}_{3n} | i12 
\rangle .
\ee
As we will see in Section \ref{lp3nf}, the form of the N$^2$LO chiral three-neutron
force is sufficiently simple that in most cases analytical formulas for the 
Landau parameters are possible.
We note that this leading-order contribution to the quasiparticle interaction from the 
N$^2$LO chiral three-neutron force is nearly equivalent to the effective 
interaction calculated previously in refs.\ 
\cite{holt09,holt10,hebeler10}, although the quasiparticle
interaction represents a restricted kinematical configuration for which the two
interacting particles lie on the Fermi surface $|\vec p_{1,2}| = k_f$.

%%%%%%%%%%%%%%%%%%%%%%%%%%%%%%%%%%%%%%%%%%%%%%%%%%%%%%%%%%%%%%%%%%%%%%%%%%%%%%%
%%%%%%%%%%%%%%%%%%%%%%%%%%%%%%%%%%%%%%%%%%%%%%%%%%%%%%%%%%%%%%%%%%%%%%%%%%%%%%%
\section{Benchmark calculations with model interactions}
\label{mi}

\begin{figure}
\begin{center}
\includegraphics[height=3cm]{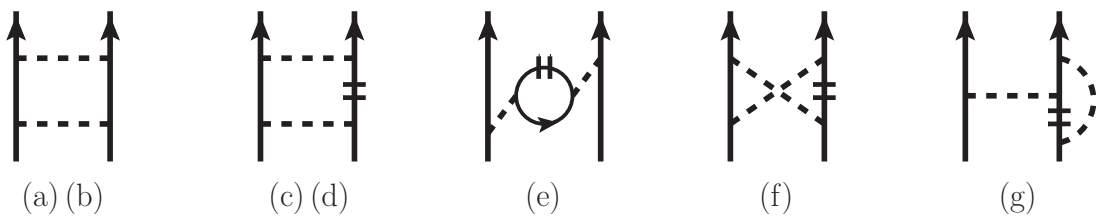}
\end{center}
\vspace{-.5cm}
\caption{Diagrammatic contributions to the second-order quasiparticle interaction in neutron
matter. Diagrams (a)--(d) correspond to the sum of particle-particle and hole-hole contributions,
while (e)--(g) together comprise the particle-hole contribution. Medium insertions are denoted by 
the short double lines, and the labels (b) and (d) refer to the crossed terms of (a) and (c).
Reflected diagrams are not shown.}
\label{qp2norbert}
\end{figure}

In order to verify the spin-decomposition techniques and the accuracy 
of the intricate numerical calculations involved in
the second-order calculation of the quasiparticle interaction, it is 
useful to examine simple model interactions that are amenable to (partial) 
analytical solutions. Intermediate-state momentum integrations and spin traces 
are carried out explicitly without decomposing the interaction into partial waves.
For details on this approach to computing Fermi liquid parameters, see ref.\ \cite{kaiser06}.
The diagrammatic contributions are shown in Fig.\ \ref{qp2norbert}, where the 
double dash on a fermion line represents a ``medium insertion''. It is defined as the 
difference between the free-space propagator and the in-medium 
propagator:
\begin{eqnarray}
&&i\left (\frac{\theta(|\vec p\,|-k_f)}{p_0-\vec p^{\,\,2}/2M_n+i\epsilon}
+\frac{\theta(k_f - |\vec p\,|)}{p_0-\vec p^{\,\,2}/2M_n-i\epsilon}\right )\nonumber \\
&&= \frac{i}{p_0-\vec p^{\,\,2}/2M_n+i\epsilon}-2\pi \delta(p_0-\vec p^{\,\,2}/2M_n)
\theta(k_f - |\vec p\,|).
\label{imp}
\end{eqnarray}
Comparing with the expression given in eq.\ (\ref{order2qp}), the sum of diagrams (a)-(d)
corresponds to ${\cal F}^{(2)}_{pp} + {\cal F}^{(2)}_{hh}$. Expanding $(1-n_3)(1-n_4) = 
1-n_3 - n_4 + n_3 n_4$ in the $pp$ diagram, we see that the term proportional to
$n_3 n_4$ cancels the hole-hole contribution. Hence, the sum of the particle-particle
and hole-hole diagrams gives just a free-space contribution and two terms with one
medium insertion. The remaining diagrams (e)-(g) in Fig.\ \ref{qp2norbert} correspond
to the particle-hole contribution in eq.\ (\ref{order2qp}).
In the following we consider only the Fermi liquid parameters $h_L(k_f),k_L(k_f),$
and $l_L(k_f)$ associated with the noncentral components of the quasiparticle interaction. 
Detailed calculations for the central components have been presented in ref.\ \cite{holt11}.
We provide explicit formulas for the relevant Landau parameters up to $L=1$ obtained 
by first decomposing the effective interaction into the relevant operators and then projecting the 
expansion coefficients onto Legendre polynomials $P_L(\hat p_1 \cdot \hat p_2)$.

\subsection{Pseudoscalar boson exchange to second order}

We begin by considering pseudoscalar boson exchange with a ``form factor'' modification:
\begin{equation} 
V(\vec q\,) = -\frac{g^2}{(m^2+q^2)^2} \, \vec \sigma_1 
\cdot \vec q\,\,\vec \sigma_2 \cdot \vec q \, ,
\label{mopem}
\end{equation}
where $g$ is a dimensionless coupling constant and $m$ is the mass parameter chosen
to be sufficiently large to achieve good convergence in momentum integrals and partial wave
summations. The momentum transfer is denoted by $\vec q$, and
the squared denominator in eq.\ (\ref{mopem}) insures that all loop integrals converge. 
As discussed previously, the leading-order (free-space) contribution to the 
quasiparticle interaction contains only the exchange tensor interaction (in addition to the
central components). Its $L=0,1$ projections
are given by
\begin{equation} 
h_0(k_f)^{(1)}  = {g^2 \over 3 m^2}
\bigg\{{1\over 4u^2}\ln(1+4u^2) -{1\over 1+4u^2} \bigg\}\, ,
\end{equation}
\begin{equation} 
h_1(k_f)^{(1)} ={g^2 \over m^2}
\bigg\{{1+u^2\over 4u^4}
\ln(1+4u^2) -{1\over u^2}+{1\over 1+4u^2}\bigg\}\, ,
\end{equation}
with the dimensionless parameter $u=k_f/m$.
At second order, the direct contributions ${\cal F}^{(2a)}$ and ${\cal F}^{(2c)}$ 
have no noncentral components. The crossed term (b) of the iterated pseudoscalar-exchange 
diagram gives a 
contribution to the exchange tensor interaction:
\begin{eqnarray} 
h_0(k_f)^{(2b)} &=& {g^4 M_n \over 48\pi
m^3}\bigg\{{1\over 2u^2}
\ln{1+2u^2 \over1+u^2} -{1\over 1+2u^2} \nonumber \\
&&+ \int_0^u\!\! dx\, {1+4x^2+8x^4 \over u^2(1+2x^2)^3}
(\arctan 2x-\arctan x)\bigg\} \,,
\label{bmp}
\end{eqnarray}
\begin{eqnarray} 
h_1(k_f)^{(2b)} &=& {g^4 M_n \over 16 \pi
m^3}\bigg\{{1\over
1+2u^2}-{1\over u^2}+{2+u^2\over 2u^4}\ln{1+2u^2\over 1+u^2} \nonumber
\\ &&
+\int_0^u\!\! dx\, {u^2-2x^2 \over u^4(1+2x^2)^3}(1+4x^2+8x^4)
(\arctan 2x-\arctan x) \bigg\} \, .
\end{eqnarray}
Iterated pseudoscalar exchange does not include medium modifications and 
therefore cannot generate an effective interaction that depends
explicitly on the center-of-mass momentum $\vec P$.
In contrast, the crossed term (d) from the planar box diagram with Pauli 
blocking gives rise to an exchange tensor force and center-of-mass
tensor force (as noted in Section \ref{qpnm}, the particle-particle 
and hole-hole diagrams cannot generate the cross-vector interaction). 
The associated $L=0,1$ exchange tensor Fermi liquid parameters are
\begin{eqnarray} 
h_0(k_f)^{(2d)} &=& {g^4 M_n \over 12\pi^2 k_f^3}
\int_0^1 \!\! dx
\int_{-1}^1 \!\! dy \int_{-1}^1 \!\! dz \,{x^2\over
[u^{-2}+A]^2[u^{-2}+B]^2}
\nonumber  \\ && \times \bigg\{y z-x(x+y+z)-{4x^2 |y+z| \over A+B-4}
+(1-x^2)^2
{(A+B) \eta \,\theta(W)\over (A+B-4) \sqrt{W}} \bigg\} \,, 
\label{bti}
\end{eqnarray}
\begin{eqnarray} 
h_1(k_f)^{(2d)} &=& {g^4 M_n \over 4\pi^2 k_f^3} \int_0^1
\!\! dx
\int_{-1}^1 \!\! dy \int_{-1}^1 \!\! dz \,{x^2\over
[u^{-2}+A]^2[u^{-2}+B]^2}
\bigg\{ x^4-x^2 yz\nonumber  \\ && +x(y+z)(2-yz)+{1\over
2}(3y^2z^2-y^2-z^2-1)
+{4x^2 |y+z| \over A+B-4}\nonumber  \\ &&  +x(1-x^2)^2(y+z-x){(A+B) \eta \,
\theta(W)\over (A+B-4) \sqrt{W}} \bigg\} \,, \end{eqnarray}
while the center-of-mass tensor Fermi liquid parameters are given by
\begin{eqnarray} 
k_0(k_f)^{(2d)} &=& {g^4 M_n \over 12\pi^2 k_f^3}
\int_0^1 \!\!
dx \int_{-1}^1 \!\! dy \int_{-1}^1 \!\! dz \,{x^2\over
[u^{-2}+A]^2[u^{-2}+B]^2}
\nonumber \\ && \times  \bigg\{y z -x(3x+y+z)-{8x^2 |y+z| \over A+B-4}
+\bigg[
2x^2(2y z-1) \nonumber  \\ && +3+3x^4-2x(1+x^2)(y+z) +{8(1-x^2)^2 \over
A+B-4}
\bigg] {\eta \,\theta(W)\over \sqrt{W}} \bigg\} \,, 
\end{eqnarray}
\begin{eqnarray} 
k_1(k_f)^{(2d)} &=& {g^4 M_n \over 4\pi^2 k_f^3}
\int_0^1 \!\! dx
\int_{-1}^1 \!\! dy \int_{-1}^1 \!\! dz \,{x^2\over
[u^{-2}+A]^2[u^{-2}+B]^2}
\bigg\{3x^4+x^2(2+y z) \nonumber  \\ && +x(y+z)(2-2x^2-y z)+{1\over
2}(3y^2z^2 -y^2-z^2-1)+{8x^2 |y+z| \over A+B-4}  \nonumber  \\ 
&& +\bigg[x(y+z)(3+4x^2y z +5x^4)-4-3x^6 -2x^4(1+y^2+4y z+z^2) \nonumber \\ 
&& +x^2(5-2y^2-4y z-2z^2) -{8(1-x^2)^2 \over A+B-4}\bigg] 
{\eta \,\theta(W)\over \sqrt{W}} \bigg\} \,,
\label{eti}
\end{eqnarray}
with abbreviations: $A = 1+x^2-2x y$, $B = 1+x^2-2x z$, $W = (x-y)^2(x-z)^2
-(1-y^2)(1-z^2)$ and $\eta= {\rm sign}[(x-y)(x-z)]$. In fact,
only the Pauli-blocked planar box diagram (d) gives rise to a center-of-mass 
tensor interaction from second-order pseudoscalar exchange.

\setlength{\tabcolsep}{.07in}
\begin{table}
\begin{tabular}{|c||D{.}{.}{1.3}|D{.}{.}{2.3}||D{.}{.}{2.3}|D{.}{.}{2.3}|} \hline
\multicolumn{5}{|c|}{Modified pseudoscalar boson exchange ($k_f=1.7$\,fm$^{-1}$)} \\ \hline
 & \multicolumn{1}{c|}{$h_0$ [fm$^2$]}
 & \multicolumn{1}{c||}{$k_0$ [fm$^2$]}
 & \multicolumn{1}{c|}{$h_1$ [fm$^2$]}
 & \multicolumn{1}{c|}{$k_1$ [fm$^2$]} \\ \hline
2nd(pp) & 0.134 & -0.142 & -0.166 &  0.053  \\ \hline
2nd(hh) & 0.027 & -0.061 & -0.057 &  0.074  \\ \hline
2nd(ph) & 1.394 & -0.001 &  0.912 & -0.001  \\ \hline \hline
Total   & 1.555 & -0.205 &  0.689 &  0.125 \\ \hline \hline
Analytical   & 1.552 & -0.209 & 0.719 & 0.126  \\ \hline
\end{tabular}
\caption{The $L=0,1$ noncentral Fermi liquid parameters from the pseudoscalar
exchange interaction in eq.\ (\ref{mopem}) at second order. We compare the sum of the 
particle-particle, hole-hole and particle-hole diagrams computed numerically to the
semi-analytical results of eqs.\ (\ref{bmp})--(\ref{emp}).}
\label{mpi}
\end{table}

Of the three diagrams (e)-(g) in Fig.\ \ref{qp2norbert} that encode the effects
of medium polarization, (e) and (g) contribute to the quasiparticle interaction 
in the crossed channel and (f) contributes in the direct channel, regardless of 
the form of two-body interaction. In the case of pseudoscalar exchange
only (e) and (g) are nonvanishing for noncentral components. Diagram (e), representing
the coupling of the boson to nucleon-hole states, yields for the exchange
tensor interaction:
\begin{equation}
h_0(k_f)^{(2e)}={8g^4 M_n \over 3\pi^2
m^3u^2}\int_0^u \!\!dx\,
{x^4 \over (1+4x^2)^4} \bigg[ 2u x +(u^2-x^2) \ln{u+x\over u-x} \bigg] \,,
\end{equation}
\begin{equation}  
h_1(k_f)^{(2e)}={8g^4 M_n \over \pi^2 m^3u^4}\int_0^u \!\!dx\,
{x^4(u^2-2x^2) \over (1+4x^2)^4} \bigg[ 2u x +(u^2-x^2) \ln{u+x\over
u-x} \bigg] \,.
\end{equation}
The density-dependent vertex correction (g) can be split into a factorizable part:
\begin{equation}  
h_0(k_f)^{(2g)} = {g^4 M_n \over 24\pi^2 m^3 u^3
}\bigg[{4u^2
\over 1+4u^2}-\ln(1+4u^2)\bigg]\bigg[{1+2u^2 \over
4u^2}\ln(1+4u^2)-1\bigg] \,,
\end{equation}
\begin{equation}  
h_1(k_f)^{(2g)} = {g^4 M_n \over 8\pi^2 m^3
u^5}\bigg[1-{1+2u^2
\over 4u^2}\ln(1+4u^2)\bigg]\bigg[(1+u^2) \ln(1+4u^2) -3u^2-{u^2\over
1+4u^2}
\bigg]\,, 
\end{equation}
and a non-factorizable part:
\begin{eqnarray} 
h_0(k_f)^{(2g')} &=& {g^4 M_n \over 6\pi^2 m^3 u^2}
\int_0^u
\!\! dx \, \bigg[\ln(1+4x^2)-{4x^2 \over 1+ 4x^2}\bigg]\bigg\{ {2ux
(1+4u^2)^{-1}
\over 1+4u^2-4x^2} \nonumber  \\ && +{u^2-x^2\over  (1+4u^2-4x^2)^{3/2}}
\ln {(u\sqrt{1+4u^2-4x^2} +x )^2 \over(1+4u^2)(u^2-x^2)} \bigg\} \,,
\end{eqnarray}
\begin{eqnarray}  
h_1(k_f)^{(2g')} &=& {g^4 M_n \over 8\pi^2 m^3 u^4}
\int_0^u
\!\! dx \, \bigg[\ln(1+4x^2)-{4x^2 \over 1+ 4x^2}\bigg] \bigg\{ {4ux
(1+2u^2)\over
(1+4u^2)(1+4u^2-4x^2)}\nonumber \\ && -\ln{u+x\over u-x}
+{1+(u^2-x^2)(6+4u^2)
\over  (1+4u^2-4x^2)^{3/2}} \ln { ( u\sqrt{1+4u^2-4x^2} +x )^2
\over(1+4u^2)(u^2-x^2)}
\bigg\} \,.
\label{emp}
\end{eqnarray}
We note that diagrams (e)--(g), representing the particle-hole contribution,
do not give rise to a cross-vector interaction in agreement with the general
argument presented in Section \ref{qpnm}.
We now evaluate the expressions given in eqs.\ (\ref{bmp})--(\ref{emp}) 
choosing $g = 5$, $m = 300$\,MeV and $k_f=1.7$\,fm$^{-1}$. In Table \ref{mpi} we compare these
semi-analytical results to those obtained from a numerical evaluation of the second-order
contributions as given in Section \ref{qpnm} through a partial-wave decomposition. The 
agreement is generally on the order of 2\% or better. Part of this discrepancy 
(in particular for the Fermi liquid parameter $h_1$) arises
from the numerical uncertainty in the triple integrations of eqs.\ (\ref{bti})-(\ref{eti}).

\subsection{Spin-orbit interaction to second order}

We consider as well the case of a pure spin-orbit interaction
of the form
\begin{equation} 
V_{\rm so} = {2g_s^2 \over (m_s^2+q^2)^2} \, i \, (\vec
\sigma_1 +\vec \sigma_2) \cdot (\vec q \times \vec p\,)\,,
\label{soi}
\end{equation}
where $\vec q$ is the momentum transfer, $\vec p$ is half the incoming relative
momentum and $m_s$ is the mass of the exchanged boson to be fixed later.
We consider only the isotropic ($L=0$) contributions to the noncentral
interactions. The first-order contribution from $V_{so}$ to the quasiparticle
interaction vanishes trivially since $\vec q \times \vec p = 0$. 
The direct term (a) of the planar box diagram gives the contribution
\begin{equation}
h_0(k_f)^{(2a)} = {g_s^4 M_n \over 288\pi m_s^3}
\bigg\{{5+12u^2
\over 1+4u^2}-{5\over 4u^2} \ln(1+4u^2)\bigg\}\,, 
\label{bso}
\end{equation}
with $u=k_f/m_s$, while the crossed term (b) contribution reads
\begin{equation} 
h_0(k_f)^{(2b)} = {g_s^4 M_n \over 24\pi
m_s^3}\bigg\{{1\over 2u^2}
\ln{1+2u^2 \over1+u^2} -{1\over 1+2u^2} +\int_0^u\!\! dx\,
{1+4x^2+8x^4 \over u^2(1+2x^2)^3}(\arctan 2x-\arctan x)\bigg\} \,.
\end{equation}
For the Pauli-blocked planar box diagram, we find that the direct
term (c) gives rise to both exchange tensor as well as center-of-mass
tensor contributions of the form
\begin{eqnarray} 
h_0(k_f)^{(2c)} &=& {g_s^4 M_n \over 6\pi^2 k_f^3}
\int_0^1 \!\! dx
\int_{-1}^1 \!\! dy \int_{-1}^1 \!\! dz \,{x^2\over [u^{-2}+A]^4} 
\bigg\{-x(x+y)
\nonumber  \\ && -{4x^2 |y+z| \over A+B-4} +(1-x^2)^2{(A+B) \eta \,\theta(W)
\over (A+B-4) \sqrt{W}} \bigg\} \,
\end{eqnarray}
\begin{eqnarray} 
k_0(k_f)^{(2c)} &=& {g_s^4 M_n \over 6\pi^2 k_f^3} \int_0^1 \!\!
dx \int_{-1}^1 \!\! dy \int_{-1}^1 \!\! dz \,{x^2\over [u^{-2}+A]^4} \bigg\{
-x(3x+y)-{8x^2 |y+z| \over A+B-4}  \nonumber  \\ &&+\bigg[3+3x^4+2x^2(2y
z-1) -2x(1+x^2)(y+z) +{8(1-x^2)^2 \over A+B-4}\bigg] {\eta \,\theta(W)\over
\sqrt{W}} \bigg\} \,, 
\end{eqnarray}
with abbreviations: $A = 1+x^2-2x y$, $B = 1+x^2-2x z$, $W = (x-y)^2(x-z)^2
-(1-y^2)(1-z^2)$ and $\eta= {\rm sign}[(x-y)(x-z)]$.
Likewise, the crossed term (d) of the Pauli-blocked planar box diagram yields
\begin{eqnarray} 
h_0(k_f)^{(2d)} &=& {g_s^4 M_n \over 6\pi^2 k_f^3} \int_0^1 \!\! dx
\int_{-1}^1 \!\! dy \int_{-1}^1 \!\! dz \,{x^2\over [u^{-2}+A]^2[u^{-2}+B]^2}
\nonumber  \\ 
&& \times \bigg\{y z-x(x+y+z)-{4x^2 |y+z| \over A+B-4} +(1-x^2)^2
{(A+B) \eta \,\theta(W)\over (A+B-4) \sqrt{W}} \bigg\} \,,
\end{eqnarray}
\begin{eqnarray} 
&& k_0(k_f)^{(2d)} = {g_s^4 M_n \over 6\pi^2 k_f^3} \int_0^1 \!\!
dx \int_{-1}^1 \!\! dy \int_{-1}^1 \!\! dz \,{x^2\over
[u^{-2}+A]^2[u^{-2}+B]^2} \bigg\{y z -x(3x+y+z) \nonumber \\ 
&& -{8x^2 |y+z| \over A+B-4} +\bigg[
2x^2(2y z-1) +3+3x^4-2x(1+x^2)(y+z) +{8(1-x^2)^2 \over
A+B-4} \bigg] {\eta \,\theta(W)\over \sqrt{W}} \bigg\} \,.
\end{eqnarray}

\setlength{\tabcolsep}{.07in}
\begin{table}
\begin{tabular}{|c||D{.}{.}{2.3}|D{.}{.}{2.3}|} \hline
\multicolumn{3}{|c|}{Spin-orbit interaction ($k_f=1.7$ fm$^{-1}$)} \\ \hline
 & \multicolumn{1}{c|}{$h_0$ [fm$^2$]} 
 & \multicolumn{1}{c|}{$k_0$ [fm$^2$]} \\ \hline 
2nd(pp+hh) & 0.761 & -0.311  \\ \hline
2nd(ph) & -0.758 & -0.444   \\ \hline 
Total   & 0.003 & -0.756   \\ \hline \hline
Analytical (pp+hh)  & 0.772 & -0.312   \\ \hline
Analytical (ph)  & -0.761 & -0.448   \\ \hline
Total  & 0.011 & -0.760   \\ \hline
\end{tabular}
\caption{The $L=0$ noncentral Fermi liquid parameters for the spin-orbit 
interaction at second-order. The numerical results for $h_0$ and $k_0$ based on 
the partial-wave 
decomposition are compared with values from the semi-analytical formulas in eqs.\ 
(\ref{bso})--(\ref{eso}).}
\label{mso}
\end{table}

The coupling to intermediate particle-hole states (e) in the crossed channel 
generates both exchange and center-of-mass tensor interactions:
\begin{equation} 
h_0(k_f)^{(2e)} = {g_s^4 M_n\over (3\pi
u)^2m_s^3}\int_0^u\!\! dx\,
{x^2\over (1+4x^2)^4}\bigg[u x(15x^2-17u^2)-{15\over 2}(u^2-x^2)^2 \ln{u+x
\over u-x}\bigg]\,, \end{equation}
\begin{equation}  k_0(k_f)^{(2e)} = {2g_s^4 M_n\over 3\pi^2 u^2
m_s^3}\int_0^u\!\!
dx\, {x^2(x^2-u^2) \over (1+4x^2)^4} \bigg[2u x+(u^2-x^2)\ln{u+x \over u-x}
\bigg]\,,
\end{equation}
and for the Pauli-blocked crossed box diagram (f), the direct term produces the
contributions
\begin{eqnarray}
h_0(k_f)^{(2f)} &=& {8g_s^4 M_n \over (3\pi u)^2
m_s^3}\int_0^u\!\!
dx\,{x^3 \over (1+4x^2)^4} \bigg[u(u^2-x^2)+2u^3
\ln{u^2-x^2 \over 4u^2} \nonumber  \\ &&-6u x^2\ln{u^2-x^2 \over 4x^2}-4x^3 
\ln{u+x \over u-x}\bigg] \,, 
\end{eqnarray}
\begin{eqnarray} 
k_0(k_f)^{(2f)} &=& {8g_s^4 M_n \over 3\pi^2 u^2 m_s^3}\int_0^u
\!\! dx\,{x^3 \over (1+4x^2)^4}\bigg[u(u^2-x^2)-4u x^2
\ln{u^2-x^2 \over 4x^2}\nonumber  \\ &&-2x(u^2+x^2) \ln{u+x \over
u-x}\bigg] \,.
\end{eqnarray}
Finally, the crossed term of the vertex correction diagram (g) reads
\begin{eqnarray} 
h_0(k_f)^{(2g)} &=& {64g_s^4 M_n \over 3\pi^2 k_f^3} \int_0^u
\!\!dx\int_0^u \!\! dy \,{x^3 y^2\over (1+4x^2)^2 (1+4y^2)^2} \bigg\{u\,
{\rm Re}
\ln{x+\sqrt{x^2+y^2-u^2}\over u+y} \nonumber  \\ &&+{y\over u^2}(u^2-x^2)
+{x \over u^2-y^2} \Big[ u\, {\rm Re}\sqrt{x^2+y^2-u^2}-x y \Big]
\bigg\} \,, 
\end{eqnarray}
\begin{eqnarray} 
k_0(k_f)^{(2g)} &=& {64g_s^4 M_n \over 3\pi^2 k_f^3} \int_0^u
\!\!dx\int_0^u \!\! dy \,{x^3 y^2\over (1+4x^2)^2 (1+4y^2)^2}  \nonumber
\\ &&  \times \bigg\{{y\over u^2}(u^2-x^2) +{2x \over u^2-y^2} \Big[ u\,
{\rm Re} \sqrt{x^2+y^2-u^2} -x y \Big] \bigg\} \,,
\label{eso}
\end{eqnarray}
where Re stands for real part. Again, we find that the spin-orbit interaction iterated
to second order does not give rise to a cross-vector interaction.
We choose the parameters $g_s=10$, $m_s=700$\,MeV and $k_f=1.7$\,fm$^{-1}$. 
In Table \ref{mso} we compare the 
sum of the particle-particle and hole-hole diagrams to the sum of diagrams
(a)--(d) in Fig.\ \ref{qp2norbert}. Likewise we compare the particle-hole
contribution to the sum of diagrams (e)--(g) in Fig.\ \ref{qp2norbert}.
The results of both calculations are again in very good numerical agreement with one another.

\subsection{Spin-nonconserving cross-vector interaction}

As mentioned in subsection \ref{soc}, the origin of the 
spin-nonconserving quasiparticle interaction (proportional to the cross-vector
operator $(\sigma_1 \times \sigma_2)\cdot (\hat q \times \hat P)$) is the
interference of the spin-orbit component of the two-body potential with its other 
(central, spin-spin and tensor) components. In order to exemplify this mechanism
through a solvable model, we consider a simple contact interaction with couplings
in all relevant channels:
\begin{equation} V_{\rm ct} = \Gamma_c + \Gamma_s\, \vec\sigma_1\cdot
\vec\sigma_2 +
  \Gamma_t\, \vec\sigma_1\cdot\vec q\,\, \vec\sigma_2 \cdot\vec q + i
\Gamma_{so}
(\vec\sigma_1+ \vec\sigma_2) \cdot (\vec q \times \vec p\,)\,. \end{equation}
At second order the interference terms arising from the particle-hole
diagrams (e)--(g) in Fig.\ \ref{qp2norbert} can be worked out analytically. One finds for the first three 
Landau parameters of the cross-vector interaction
\begin{equation} l_0 = {M_n k_f^3 \over 5 \pi }(\Gamma_c-3\Gamma_s)
\Gamma_{so} \,, \quad  l_1 = {3M_nk_f^3 \over 70 \pi }(\Gamma_c-3\Gamma_s)
\Gamma_{so} \,, \quad l_2 = {11M_nk_f^3 \over 84 \pi }(3\Gamma_s-\Gamma_c)
\Gamma_{so} \,,\end{equation}
\begin{equation} l_0 = -{8M_nk_f^5 \over 21 \pi }\Gamma_t \Gamma_{so}\,,
\qquad
  l_1 = -{4M_nk_f^5 \over 21 \pi }\Gamma_t \Gamma_{so} \,, \qquad l_2 =
{50M_nk_f^5 \over 231 \pi }\Gamma_t \Gamma_{so}\,.
\end{equation}
Due to their different structure we have listed separately the
interference terms with
central and spin-spin interactions and the ``tensor-type'' interaction
$\Gamma_t\,\vec \sigma_1 \cdot \vec q \, \vec \sigma_2 \cdot \vec q$. 
Our second-order calculation based on
a decomposition of the two-body potential into partial-wave matrix elements
reproduced these
analytical results with good numerical accuracy. In the absence of the tensor
term, the condition $\Gamma_c=3\Gamma_s$ (giving $l_L=0$) is equivalent to a vanishing
interaction in the spin-singlet state.
\begin{figure}
\begin{center}
\includegraphics[scale=0.9,clip]{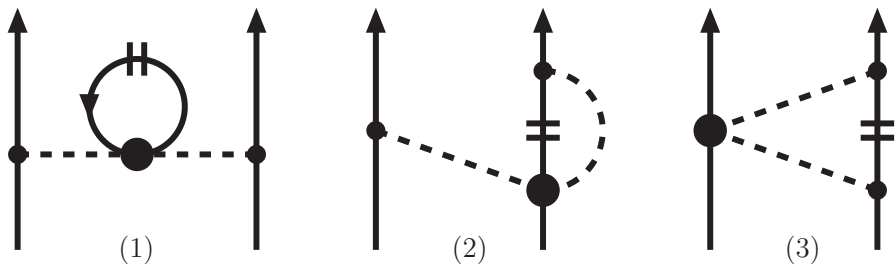}
\end{center}
\vspace{-.5cm}
\caption{Diagrammatic contributions to the quasiparticle interaction in 
neutron matter generated from the two-pion exchange three-neutron interaction. 
The short double-line symbolizes summation over the filled Fermi sea of neutrons. 
Reflected diagrams of (2) and (3) are not shown.}
\label{mfig1}
\end{figure}

%%%%%%%%%%%%%%%%%%%%%%%%%%%%%%%%%%%%%%%%%%%%%%%%%%%%%%%%%%%%%%%%%%%%%%%%%%%%%%%
%%%%%%%%%%%%%%%%%%%%%%%%%%%%%%%%%%%%%%%%%%%%%%%%%%%%%%%%%%%%%%%%%%%%%%%%%%%%%%%
\section{Landau parameters from chiral 3N interaction}
\label{lp3nf}

In this section we consider the N$^2$LO chiral three-nucleon force 
in neutron matter and derive expressions for all $L=0,1$ Landau parameters 
arising from the leading-order (one-loop) contribution to the quasiparticle interaction. 
In most cases it is possible to obtain analytical expressions for arbitrary values
of $L$, but for brevity we show only the formulas for the isotropic ($L=0$) and $p$-wave
($L=1$) parameters. As mentioned 
in subsection \ref{ctnf}, we keep only the two-pion exchange three-neutron force 
proportional to $c_1$ and $c_3$. There are three one-loop diagrams contributing
to the effective interaction, shown in Fig.\ \ref{mfig1}, which we label as 
$V_{NN}^{{\rm med},i}$ for $i=1,2,3$. 
Both the pion self-energy correction $V_{NN}^{\rm med,1}$ and the vertex correction
$V_{NN}^{\rm med,2}$ produce an effective interaction similar to that of one-pion 
exchange. Both terms vanish in the direct channel
but give contributions to $f,g$ and $h$ in the exchange channel. For the pion
self-energy correction we find
\begin{equation}  {\cal F}_0(k_f)^{(\rm med,1)} = (3-\vec \sigma_1\cdot \vec
\sigma_2+2S_{12}(\hat q)){g_A^2 m_\pi^3  \over (6\pi)^2 f_\pi^4} \bigg\{ {(2c_1-c_3)u^3
\over
1+4u^2} -c_3 u^3  +(c_3-c_1) {u \over 2} \ln(1+4u^2) \bigg\}\,,
\label{tnfb}
\end{equation}
\begin{eqnarray}  {\cal F}_1(k_f)^{(\rm med,1)} &=& (3-\vec
\sigma_1\cdot \vec \sigma_2+2S_{12}(\hat q)) {g_A^2 m_\pi^3\over 48\pi^2 f_\pi^4} \bigg\{ {(2c_1-c_3)u
\over 1+4u^2}
+(6c_1-5c_3)u \nonumber \\ &&  +\Big[ 2(c_3-c_1)u +{3c_3-4c_1 \over
2u}\Big]
\ln(1+4u^2) \bigg\}\,,
\end{eqnarray}
with $u =k_f/m_\pi$. As seen from the above formulas, the expressions for
the Fermi liquid parameters of the central, spin-spin and tensor quasiparticle
interaction are identical up to integer factors characteristic of a one-pion
exchange nucleon-nucleon potential.
The contribution from the crossed term of the pion exchange vertex
correction $V_{NN}^{\rm med,2}$ reads
\begin{eqnarray}   {\cal F}_0(k_f)^{(\rm med,2)} &=& (3-\vec
\sigma_1\cdot \vec \sigma_2+2S_{12}(\hat q)) 
{g_A^2 m_\pi^3 \over (24\pi)^2 f_\pi^4} \bigg\{{3c_1 \over 4u^5}
\Big[8u^4+4u^2-(1+4u^2) \ln(1+4u^2) \Big]\nonumber \\ && \times
\Big[ 4u^2- \ln(1+4u^2)\Big] + c_3 \bigg[ {4\over u^2} \Big(4u^2-
\ln(1+4u^2)
\Big) \arctan 2u \nonumber \\ &&  +{48u^4+16u^2+3 \over 32u^7} 
\ln^2(1+4u^2)
+ {40u^3 \over 3} -22 u +{2\over u} +{3 \over 2u^3}\nonumber \\ && +
{12u^4-16u^6-30u^2-9 \over 12u^5} \ln(1+4u^2) \bigg] \bigg\}\,,
\end{eqnarray}
\begin{eqnarray}  &&{\cal F}_1(k_f)^{(\rm med,2)} = (3-\vec
\sigma_1\cdot \vec \sigma_2+2S_{12}(\hat q)) {g_A^2 m_\pi^3  \over (16\pi)^2 
f_\pi^4} \bigg\{{c_1 \over 2u^7}\Big[ 4u^2- (1+2u^2)\ln(1+4u^2)\Big]
\nonumber \\ 
&& \times \Big[8u^4+4u^2-(1+4u^2) \ln(1+4u^2)\Big] +{c_3\over 3u^4}\bigg[ 8 \Big(4u^2-
(1+2u^2)
\ln(1+4u^2)\Big) \arctan 2u \nonumber \\ && +{96u^6+80u^4+22u^2+3 \over
16u^5} \ln^2(1+4u^2) + {56u^6 -32u^8-60u^4-48u^2-9 \over 6u^3}
\ln(1+4u^2) \nonumber \\ &&  +{8u^5 \over 3}(7-4u^2) -28 u^3 +10u +{3
\over u}
\bigg] \bigg\}\,.
\end{eqnarray}

Lastly, we compute the $L=0,1$ Landau parameters associated with the 
Pauli-blocked two-pion exchange diagram $V_{NN}^{\rm med,3}$, which has
a more complicated spin structure than $V_{NN}^{\rm med,1}$ and
$V_{NN}^{\rm med,2}$. Both the direct and exchange terms contribute,
and the central components of the quasiparticle interaction read
\begin{eqnarray}
{\cal F}_0(k_f)^{(\rm med,3)} &=& {g_A^2 m_\pi^3 \over 16\pi^2 f_\pi^4} 
\bigg\{ 8(c_3-c_1) u - {8 c_3  u^3\over 3}  +{3c_3-4c_1 \over u}
\ln(1+4u^2) \nonumber \\ &&\hspace{-.4in}+(12 c_1-10c_3) \arctan 2u +(1+\vec
\sigma_1 \cdot \vec \sigma_2)\!\int_0^u\!\!dx \Big[ 2c_1 \, Z^2 + {c_3 \over 3}
(X^2 + 2 Y^2)  \Big] \bigg\} \,, 
\end{eqnarray}
\begin{eqnarray}
{\cal F}_1(k_f)^{(\rm med,3)} &=& (1+\vec \sigma_1\cdot \vec \sigma_2)
{g_A^2 m_\pi^3 \over 16\pi^2 f_\pi^4}\int_0^u\!\!dx  \bigg\{
2c_1(Z_a^2+2Z_b^2) + c_3 \Big(X_a^2 + 2X_b^2+{4\over 3}X_c^2 \Big) 
\bigg\} \,.
\label{tnfe}
\end{eqnarray}
The auxiliary functions $X,Y,Z$; $X_a, X_b, X_c$; and $Z_a,Z_b$ 
are defined in Section 2.2 of ref.\ \cite{holt12}.
Concerning the noncentral quasiparticle interaction, $V_{NN}^{\rm med,3}$ produces no tensor
forces in pure neutron matter. However, one expects $V_{NN}^{\rm med,3}$ to generate an 
exchange tensor force in symmetric nuclear matter \cite{holt10} since in this case
the three-nucleon force proportional to $c_4$ does not vanish. As a special feature, 
the crossed term of the Pauli-blocked $2\pi$
exchange diagram gives rise to a cross-vector interaction. 
With its usual representation given by 
\begin{equation}
{\cal F}_{\rm cross} = (\vec \sigma_1 \times \vec \sigma_2)
\cdot (\hat q \times \hat P) \sum_{L=0}^\infty
l_L(k_f) \, P_L(\hat p_1 \cdot \hat p_2) \,,
\label{cvn}
\end{equation}
a complete analytical solution for the Landau parameters $l_L(k_f)$
could not be obtained.
When choosing the alternative form of the cross vector interaction
\begin{equation} 
{\cal F}_{\rm cross} = {(\vec \sigma_1 \times \vec
\sigma_2) \cdot (\vec p_1 \times \vec p_2) \over |\vec p_1+\vec p_2|^2}
\sum_{L=0}^\infty
\tilde l_L(k_f) \, P_L(\hat p_1 \cdot \hat p_2) \,,
\end{equation}
all occuring integrals can be solved analytically in the present
case.
The results for the Landau parameters read
\begin{eqnarray} \tilde l_0(k_f)^{(\rm med,3)} &=& {g_A^2 m_\pi^3
\over(8\pi)^2
f_\pi^4} \bigg\{{c_1 \over u^3}\Big[16u^4- (1+4u^2)\ln^2(1+4u^2)\Big]
\\ && + c_3 \bigg[ 4u^3-8u -{1\over u}+{1+4u^2\over 2u^3} \ln(1+4u^2)
+\bigg( {1\over u}-{1\over 16 u^5}\bigg)\ln^2(1+4u^2)\bigg] 
\bigg\}\,,\nonumber
\end{eqnarray}
\begin{eqnarray} \tilde l_1(k_f)^{(\rm med,3)} &=& {g_A^2 m_\pi^3\over
(8\pi)^2
f_\pi^4} \bigg\{3c_1\bigg[8u - {8\over u}-{2\over u^3}+{8u^4+6u^2+1
\over u^5}
  \ln(1+4u^2) \nonumber \\ && - {32u^6+24u^4+8u^2+1 \over 8
u^7}\ln^2(1+4u^2)
\bigg] \nonumber \\ &&+ c_3 \bigg[ {20u^3\over 3}-16u -{2\over
u}-{3\over u^3}
-{3\over 4u^5} +{16u^6+28u^4+18u^2+3\over 8u^7}\nonumber \\ && \times
\ln(1+4u^2)
+{3 \over 64 u^9}(64u^8-20u^4-8u^2-1)\ln^2(1+4u^2) \bigg] \bigg\}\,.
\end{eqnarray}
The numerical results assuming the standard form, eq.\ (\ref{cvn}), of the 
cross vector interaction will be shown in the following section.

%%%%%%%%%%%%%%%%%%%%%%%%%%%%%%%%%%%%%%%%%%%%%%%%%%%%%%%%%%%%%%%%%%%%%%%%%%%%%%%

%%%%%%%%%%%%%%%%%%%%%%%%%%%%%%%%%%%%%%%%%%%%%%%%%%%%%%%%%%%%%%%%%%%%%%%%%%%%%%%

\section{Results}
\label{res}

\subsection{Neutron matter equation of state}

The Fermi liquid parameters in neutron matter, unlike those in symmetric nuclear matter
close to the saturation density, are largely unconstrained by empirical data. Given that
our perturbative treatment of the quasiparticle interaction includes only the
leading-order medium corrections from two- and three-body forces, it is useful to compare 
the associated zero temperature equation of state of neutron matter (at the same order
in perturbation theory) with those obtained using nonpertubative methods. As a benchmark
we consider variational calculations \cite{akmal98} of neutron matter employing
 the high-precision Argonne $v_{18}$ two-nucleon potential \cite{wiringa95}
together with the Urbana UIX three-body potential \cite{wiringa00}, which provide a 
realistic description of light nuclei and nuclear matter.

In the partial-wave representation of the two-body interaction the first-order 
contribution to the energy per particle $\bar E = E/A$ reads
\be
\bar E_{2n}^{(1)}(k_f)=\frac{1}{2 \pi^2 k_f^3} \sum_{lSJ} (2J+1) \int_0^{k_f} dp \,
p^2 (k_f-p)^2 (2k_f+p) \langle plSJ | \bar V | p l S J \rangle\,,
\ee
and the more intricate second-order contribution takes the form
\begin{eqnarray}
\bar E_{2n}^{(2)}(k_f) &=& \frac{6}{(4\pi)^4k_f^3}\sum_{\substack{l_1 l_2 l_3 l_4 \\ S m_s m_s^\prime J J^\prime M}}
\int_0^{2k_f} dp^\prime \, {p^\prime}^2 \int_0^{\sqrt{k_f^2-{p^\prime}^2/4}}
dp \, p^2 \int_{\sqrt{k_f^2-{p^\prime}^2/4}}^\infty dq \, q^2
N(l_1 m l_2 m^\prime l_3 m l_4 m^\prime ) \nonumber \\
&& \times i^{l_2+l_3-l_1-l_4} \frac{M_n}{p^2-q^2} {\cal C}^{JM}_{l_1m S m_s} 
{\cal C}^{JM}_{l_2m^\prime S m_s^\prime} {\cal C}^{J^\prime M}_{l_3 m S m_s} 
{\cal C}^{J^\prime M}_{l_4 m^\prime S m_s^\prime} \langle p l_1 S J | \bar V | q l_2 S J \rangle 
\langle q l_4 S J^\prime | \bar V | p l_3 S J^\prime \rangle \nonumber \\
&&\times \int_{-x_p}^{x_p} d \cos \theta_p P_{l_1}^m (\cos \theta_p)P_{l_3}^m (\cos \theta_p)
\int_{-x_q}^{x_q} d \cos \theta_q P_{l_2}^{m^\prime} (\cos \theta_q)P_{l_4}^{m^\prime} (\cos \theta_q),
\label{ea2n}
\end{eqnarray}
where $x_p = {\rm min}\{ 1,(k_f^2-p^2-{p^\prime}^2/4)/pp^\prime\}$ and
$x_q = {\rm min}\{ 1,(q^2-k_f^2+{p^\prime}^2/4)/qp^\prime\}$.
In the Appendix we provide the analytical expressions for the second-order
contributions to the energy per particle associated with the two model interactions
introduced in Section \ref{mi}.
Finally, the leading-order chiral three-neutron interaction leads to Hartree and Fock contributions
to the energy per particle of neutron matter of the combined form
\begin{eqnarray}  \bar E_{3n}^{(1)}(k_f)&=&{g_A^2 m_\pi^6\over(2\pi
f_\pi)^4}\bigg\{
(6c_1-5c_3) {u^3\over 3} \arctan 2u -{2 c_3\over 9} u^6 +(c_3-c_1)u^4
\nonumber \\ && +(3c_1-2c_3){u^2\over 6} +\bigg[{c_3\over 12}
-{c_1\over 8}+{u^2\over 4}(3c_3-4c_1)\bigg]\ln(1+4u^2)\nonumber \\ &&
+{1\over 32 u^3}\int_0^u\!\!dx \Big[6c_1 H^2+c_3 ( G_S^2 +2G_T^2)\Big]
\bigg\}\,,
  \end{eqnarray}
with $u=k_f/m_\pi$. The auxiliary functions $H$, $G_S$ and $G_T$ are defined in
eqs.\ (24--26) of ref.\ \cite{kaiser12}. These interaction contributions are
added to the relatististically improved
kinetic energy per particle $\bar E_{\rm kin}(k_f) =3k_f^2/10M_n-3k_f^4/56M_n^3$.

\begin{figure}
\begin{center}
\includegraphics[width=12cm,angle=270]{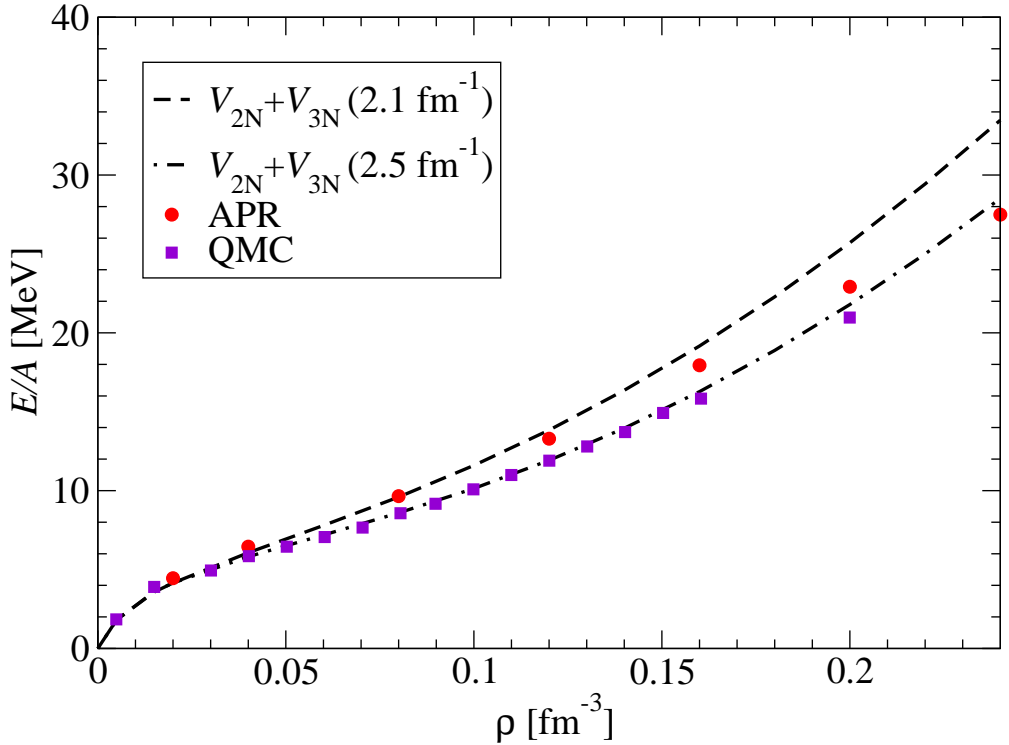}
\end{center}
\vspace{-.5cm}
\caption{Energy per particle of neutron matter from chiral and low-momentum
two- and three-body interactions. The cutoff scale associated with the bare 
chiral nuclear potential is $\Lambda = 2.5$\,fm$^{-1}$, while that of the low-momentum
interaction is $\Lambda = 2.1$\,fm$^{-1}$. The curve labeled `APR' is taken from the
variational calculations of Akmal {\it et al.} \cite{akmal98}, and the curve 
labeled `QMC' is taken from the quantum Monte Carlo calculations in refs.\
\cite{armani11,gandolfi09}. }  
\label{ea2n3n}
\end{figure}

In Fig.\ \ref{ea2n3n} we plot the resulting equation of state of neutron matter
for densities up to $\rho \simeq 1.5\rho_0$. For both chiral and low-momentum
interactions one finds good agreement with the results of the variational
calculation of ref.\ \cite{akmal98}, labeled `APR' in the figure. The difference 
between the two perturbative calculations arises 
primarily from the different values of low-energy constants $c_{1,3}$. 
Note that the recent neutron matter calculation \cite{tews12} including the subleading chiral
three- and four-neutron interactions gives very similar results.
For comparison, 
we have included in Fig.\ \ref{ea2n3n} the result for the neutron matter equation of state 
obtained in a recent quantum Monte Carlo (QMC) calculation \cite{armani11} employing 
phenomenological density-dependent two-body potentials. The data points for the lowest
two densities are taken from a different quantum Monte Carlo calculation in ref.\
\cite{gandolfi09}.

\subsection{Neutron matter quasiparticle interaction}

In this section we present and discuss the calculations of the $L=0,1,2$ Landau parameters for the
quasiparticle interaction in neutron matter from chiral and low-momentum two-
and three-nucleon forces. We begin by considering the leading-order (free-space)
contribution from both the bare chiral N$^3$LO NN interaction \cite{entem03} as well as 
the low-momentum NN potential \vlk \cite{bogner03,bogner10} obtained by integrating out momenta beyond
the cutoff scale of 2.1\,fm$^{-1}$. We use the general formula in 
eq.\ (\ref{order1qp}) for computing the first-order perturbative contribution
to the quasiparticle interaction as well as the 
projection formulas in eq.\ (\ref{projform}) for extracting the scalar functions $f,g,$ and $h$. 
In Table \ref{2nf} we show the results
for neutron matter with a Fermi momentum of $k_f=1.7$\,fm$^{-1}$ corresponding to a density of $\rho_0 = 
0.166$\,fm$^{-3}$. In both cases the Fermi liquid parameters decrease
rapidly in magnitude with $L$ for all channels. For larger values of $L$ the
difference between bare and evolved two-neutron interactions is strongly reduced,
 and at $L=3$ it is almost negligible. Short-distance repulsion in the bare
chiral NN interaction is integrated out through the renormalization-group evolution,
leading to a significant reduction in $f_0$. For both potentials, 
the compressibility ${\cal K}$ of neutron matter at $\rho_0$ would be
unphysically small \cite{fantoni01}. The isotropic component of the spin-spin interaction, $g_0$,
is enhanced by $\sim 15$\% at the resolution scale of 2.1\,fm$^{-1}$, while the 
isotropic component $h_0$ of the exchange tensor interaction is reduced by a slightly smaller factor.
The $p$-wave component of the spin-independent quasiparticle interaction, $f_1$,
increases as the resolution scale decreases. The corresponding values of the 
quasiparticle effective mass are $M^*/M_n = 0.78$ and 0.84 for the chiral and 
low-momentum interactions respectively.

Next we compute the second-order particle-particle, hole-hole and particle-hole diagrams,
shown in Fig.\ \ref{pphhph}, with two-neutron interactions. These provide the leading-order Pauli-blocking and 
polarization effects and give rise to components of the quasiparticle interaction
depending explicitly on the center-of-mass momentum $\vec P = \vec p_1 + \vec p_2$. In Table \ref{2nf2nd} we display 
the Fermi liquid parameters associated with the second-order contributions ($pp$, $hh$ and
$ph$)
for both the chiral N$^3$LO nucleon-nucleon interaction and the low-momentum
interaction \vlk in neutron matter at a density $\rho_0$. 
For the isotropic components, it is generally true that the particle-particle
diagram gives contributions that are significantly larger than the hole-hole
diagram. However, for higher values of $L$ this is not the case, and the hole-hole
diagram is in general comparable in magnitude to the particle-particle contribution.
Coherent effects among the three diagrams are observed especially
for the Landau parameters $f_1$, $g_0$, $h_0$ and $k_0$. 
Second-order effects thus tend to dramatically increase the quasiparticle
effective mass $M^*$, decrease the spin susceptibility of neutron matter and reduce
the isotropic exchange tensor strength $h_0$. Although individually large,
the contributions to $f_0$ nearly cancel at second order for the bare chiral NN
interaction. With the low-momentum interaction the sum gives $f_0 = 0.379$\,fm$^2$, which 
approximately cancels the reduction in $f_0$ from the renormalization group evolution at first order.
In general, the particle-particle contributions decrease significantly in magnitude 
as the resolution scale is lowered. This effect is due primarily to the reduction in phase space
in the particle-particle channel as the momentum cutoff is lowered.
The results compiled in Table \ref{2nf2nd} were obtained with free particle energies in the denominators
of eq.\ (\ref{order2qp}). In all subsequent tables, we include as well the one-loop
corrections to the single-particle energies in the second-order diagrams
(for details see ref.\ \cite{holt11}).

\begin{table}
\begin{tabular}{|c||D{.}{.}{2.3}|D{.}{.}{2.3}|D{.}{.}{2.3}|D{.}{.}{2.3}||D{.}{.}{2.3}|D{.}{.}{2.3}|D{.}{.}{2.3}|D{.}{.}{2.3}|}\hline
\multicolumn{1}{|c||}{$k_f=1.7$\,fm$^{-1}$} & \multicolumn{4}{c||}{Chiral N$^3$LO }
& \multicolumn{4}{c|}{$V_{\rm low-k}^{(2.1)}$} \\ \hline
\multicolumn{1}{|c||}{$L$} & \multicolumn{1}{c|}{0} & \multicolumn{1}{c|}{1} & 
\multicolumn{1}{c|}{2} & \multicolumn{1}{c||}{3} & \multicolumn{1}{c|}{0} & 
\multicolumn{1}{c|}{1} & \multicolumn{1}{c|}{2} & \multicolumn{1}{c|}{3} \\ \hline 
$f$ [fm$^2$] & -0.700 & -1.025 & -0.230 & -0.112
& -1.188 & -0.679 & -0.298 & -0.110 \\ \hline 
$g$ [fm$^2$] & 1.053 & 0.613 & 0.337 & 0.197 
& 1.212 & 0.654 & 0.346 & 0.195  \\ \hline 
$h$ [fm$^2$] & 0.270 & 0.060 & -0.040 & -0.080
& 0.239 & 0.102 & -0.051 & -0.079\\ \hline
\end{tabular}
\caption{The $L=0,1,2,3$ Fermi liquid parameters of the bare N$^3$LO chiral NN potential
of ref.\ \cite{entem03} as well as the low-momentum NN potential \vlk at a resolution
scale of 2.1\,fm$^{-1}$ at first order in many-body perturbation theory.}
\label{2nf}
\end{table}

\begin{table}
\begin{tabular}{|c|c||D{.}{.}{2.3}|D{.}{.}{2.3}|D{.}{.}{2.3}||D{.}{.}{2.3}|D{.}{.}{2.3}|D{.}{.}{2.3}|}\hline
\multicolumn{2}{|c||}{$k_f=1.7$\,fm$^{-1}$} & \multicolumn{3}{c||}{Chiral N$^3$LO}
& \multicolumn{3}{c|}{$V_{\rm low-k}^{2.1}$} \\ \hline
\multicolumn{2}{|c||}{$L$} & \multicolumn{1}{c|}{0} & \multicolumn{1}{c|}{1} & 
\multicolumn{1}{c||}{2}  & \multicolumn{1}{c|}{0} & 
\multicolumn{1}{c|}{1} & \multicolumn{1}{c|}{2}  \\ \hline 
$f$ [fm$^2$] & pp & -0.773 & 0.547 & -0.290 
& -0.225 & 0.042 & -0.124  \\ %\hline
$f$ [fm$^2$] & hh & -0.151 & 0.168 & -0.121 
& -0.161 & 0.133 & -0.063  \\ %\hline
$f$ [fm$^2$] & ph &  0.993 & 0.482 & -0.089 
& 0.765 & 0.795 & 0.255  \\ \hline 
\multicolumn{2}{|c||}{Total} & 0.069 & 1.197 & -0.500 
& 0.379 & 0.970 & 0.068  \\ \hline \hline
$g$ [fm$^2$] & pp & 0.225 & 0.086 & 0.072  
& 0.030 & 0.127 & 0.062  \\ %\hline
$g$ [fm$^2$] & hh & 0.006 & 0.089 & -0.057 
& 0.045 & 0.062 & -0.059  \\ %\hline
$g$ [fm$^2$] & ph & 0.061 & -0.016 & -0.103  
& 0.020 & -0.011 & -0.044  \\ \hline 
\multicolumn{2}{|c||}{Total} & 0.293 & 0.159 & -0.089
& 0.094 & 0.178 & -0.040  \\ \hline \hline
$h$ [fm$^2$] & pp & -0.101 & 0.112 & -0.028 
& -0.047 & 0.042 & -0.004 \\ %\hline
$h$ [fm$^2$] & hh & -0.049 & 0.084 & -0.049 
& -0.037 & 0.062 & -0.032 \\ %\hline
$h$ [fm$^2$] & ph & -0.062 & -0.090 & -0.066
& -0.108 & -0.116 & -0.044 \\ \hline 
\multicolumn{2}{|c||}{Total} & -0.212 & 0.106 & -0.143 
& -0.192 & -0.012 & -0.080  \\ \hline \hline
$k$ [fm$^2$] & pp & -0.085 & 0.064 & 0.014 
& -0.057 & 0.037 & 0.010 \\ %\hline
$k$ [fm$^2$] & hh & -0.036 & 0.052 & -0.017 
& -0.028 & 0.039 & -0.008 \\ %\hline
$k$ [fm$^2$] & ph & -0.058 & -0.017 & 0.075 
& -0.034 & -0.019 & 0.042 \\ \hline 
\multicolumn{2}{|c||}{Total} & -0.178 & 0.098 & 0.072 
& -0.119 & 0.056 & 0.043  \\ \hline \hline
%$l$ [fm$^2$] & pp & 0&270 & 0&060 & -0&040 
%& 0&239 & 0&102 & -0&051 \\ \hline
%$l$ [fm$^2$] & hh & 0&270 & 0&060 & -0&040 
%& 0&239 & 0&102 & -0&051 \\ \hline
$l$ [fm$^2$] & ph & 0.135 & -0.031 & -0.279 
& -0.062 & -0.147 & -0.161 \\ \hline 
%\multicolumn{2}{|c||}{Total} & -0&700 & -1&025 & -0&230 
%& -1&188 & -0&679 & -0&298  \\ \hline \hline
\end{tabular}
\caption{Second-order contributions to the $L=0,1,2$ Fermi liquid parameters 
in neutron matter characterized by the Fermi momentum $k_f = 1.7$\,fm$^{-1}$. 
We have separately listed the particle-particle (pp), hole-hole (hh)
and particle-hole (ph) contributions for both the bare N$^3$LO chiral NN interaction
as well as the low-momentum NN potential \vlk with $\Lambda_{\rm low-k}=2.1$\,fm$^{-1}$.}
\label{2nf2nd}
\end{table}

\begin{figure}
\begin{center}
\includegraphics[width=12cm,angle=270]{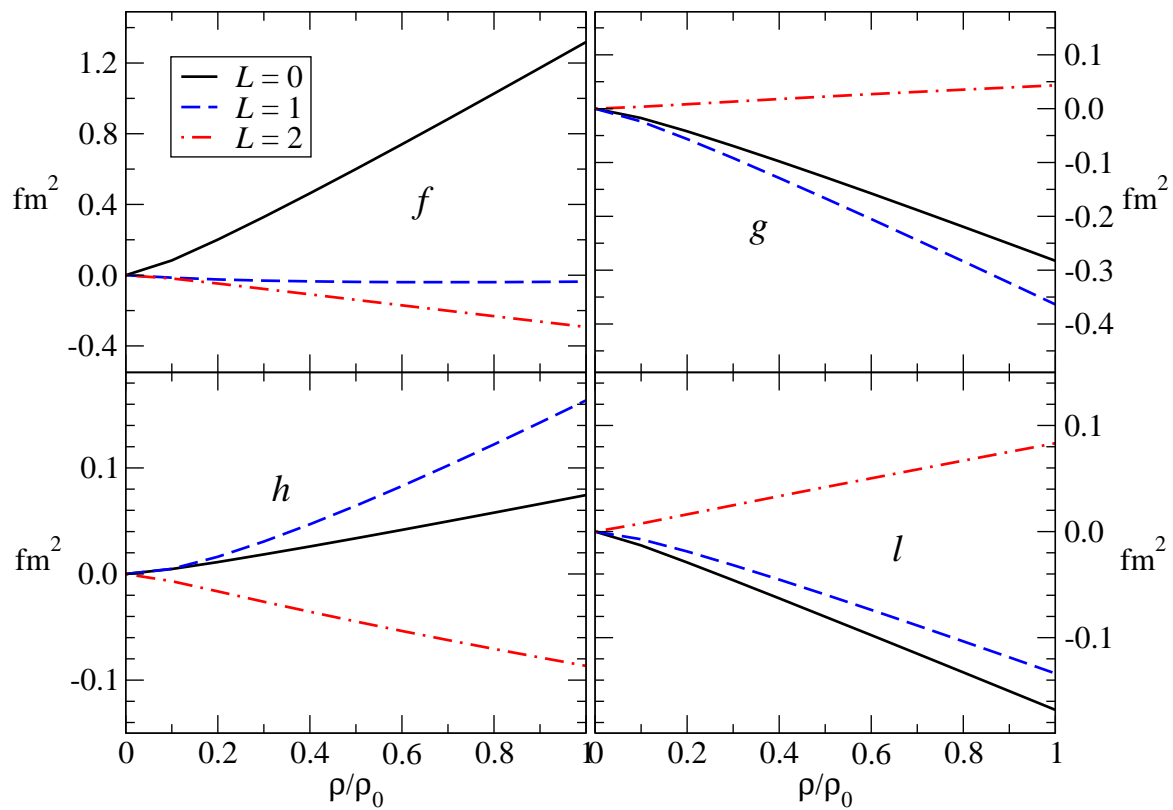}
\end{center}
\vspace{-.5cm}
\caption{Density-dependent Fermi liquid parameters from the N$^2$LO chiral 
three-nucleon force for the quasiparticle interaction in neutron matter.
The low energy constants have the values $c_1=-0.81$\,GeV$^{-1}$ and 
$c_3 = -3.2$\,GeV$^{-1}$.}
\label{dd3nflp}
\end{figure}

\begin{figure}
\begin{center}
\includegraphics[width=12cm,angle=270]{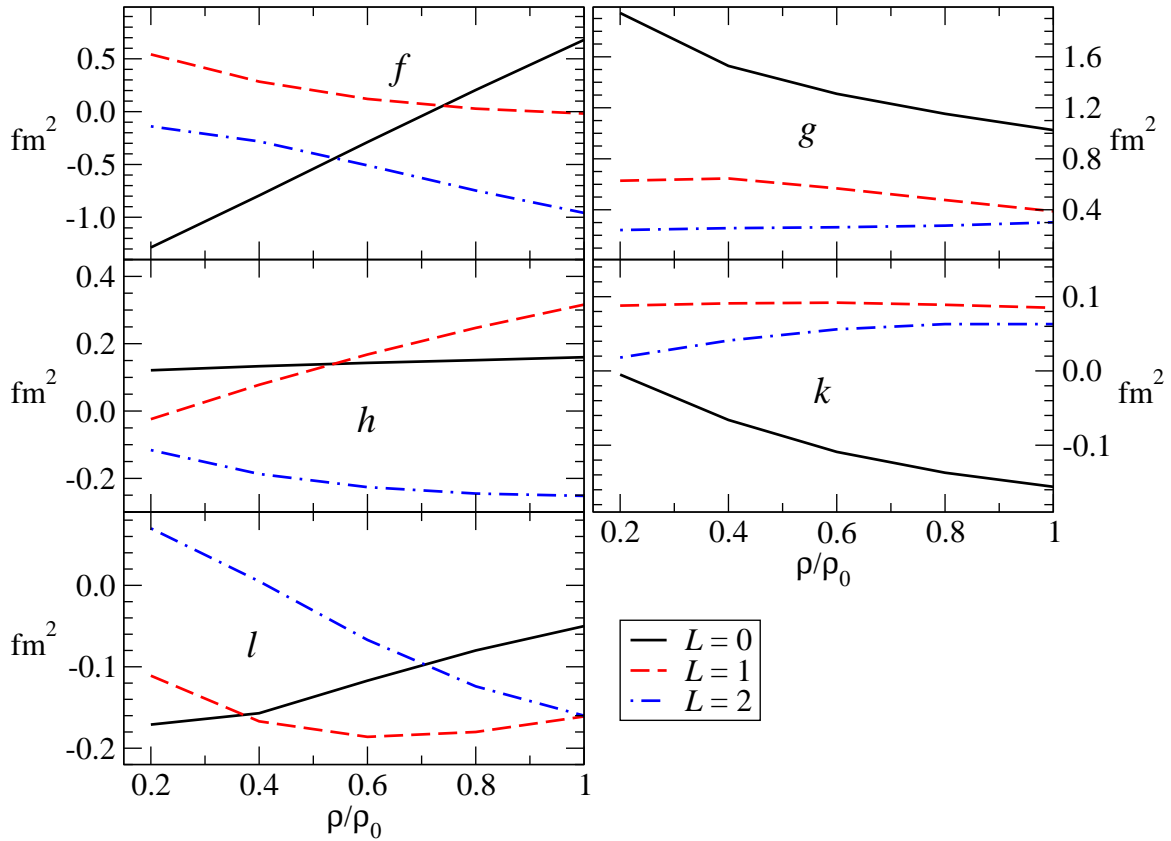}
\end{center}
\vspace{-.5cm}
\caption{Density-dependent Fermi liquid parameters inlcuding first- and second-order
contributions from the chiral N$^3$LO nucleon-nucleon potential of ref.\ \cite{entem03}
as well as the N$^2$LO chiral three-nucleon force to leading order.}
\label{ddflp}
\end{figure}

Finally, we calculate the contributions to the Fermi liquid parameters from
the leading-order chiral three-neutron force. We evaluate 
the analytical formulas in Eqs.\ (\ref{tnfb})--(\ref{tnfe}) for the central
and exchange tensor contributions and perform numerical calculations of the
cross-vector Fermi liquid parameters $l_L(k_f)$ in the standard representation, eq.\ 
(\ref{cvn}).
In Fig.\ \ref{ddflp} we plot the density-dependent Fermi liquid parameters
for the chiral three-neutron force with low-energy constants
$c_1=-0.81$\, GeV$^{-1}$ and $c_3 = -3.2$\,GeV$^{-1}$. For densities larger
than $\rho \simeq 0.5\rho_0$ the Landau parameters scale approximately linearly
with the density. The largest effect is a strong
additional repulsion in the isotropic spin-independent parameter $f_0$. In fact,
at nuclear matter saturation density $\rho_0$, the strength of the three-body correction
in this channel is larger than
the two-neutron force contributions at 1st and 2nd order together. The quasiparticle effective
mass $M^*$, governed by the parameter $f_1$, is reduced by less than 5\% at saturation density
$\rho_0$ with the
inclusion of the two-pion exchange three-neutron force. Similar
observations have been made in ref.\ \cite{holt12} for the case of symmetric nuclear matter where
additional three-nucleon forces proportional to the low-energy constants $c_4$, $c_D$
and $c_E$ are present. The qualitative similarities between the quasiparticle
interaction in neutron matter and symmetric nuclear matter from three-body forces
are due to the dominant role played by contributions proportional to the low-energy
constant $c_3$ \cite{holt12}.
Neglecting effects from noncentral Fermi liquid parameters, the chiral three-neutron force at
leading order tends to enhance the spin susceptibility $\chi$ of neutron matter 
(see eq.\ (\ref{susc})) by lowering
$g_0$ by about 25\% at $\rho_0$ from the value obtained with two-nucleon forces only. The
three-body correction to the exchange tensor interaction is smaller in
magnitude than the central force contributions. For the exchange tensor
interaction the 
pion-self energy correction $V_{NN}^{\rm med,1}$ and 
vertex correction $V_{NN}^{\rm med,2}$ enter. As seen already in ref.\
\cite{holt12,holt10} both of these terms have the structure of one-pion exchange and
are individually large with opposite sign. 
The cross-vector interaction from chiral three-nucleon forces is relatively
small compared with the central quasiparticle interactions. However, at saturation density $\rho_0$ the 
magnitudes of the cross-vector Fermi liquid parameters $l_{0,1,2}$ are comparable in magnitude to
those from two-body potential at second order.

In Fig.\ \ref{ddflp} we combine the results for the first- and second-order two-body
contributions with the leading-order three-body corrections at a resolution scale of 
$\Lambda = 2.5$\,fm$^{-1}$. All Fermi liquid 
parameters up to $L=2$ are plotted as a function of the neutron density $\rho$.
The same quantities are compiled in Table \ref{flpkf} at a Fermi momentum of $k_f = 1.7$\,fm$^{-1}$ together with
the corresponding values from the low-momentum interaction \vlk at the scale $\Lambda =2.1$\,fm$^{-1}$. 
From Fig.\ \ref{ddflp} one observes that all of the Fermi liquid parameters 
considered here for the spin-independent part of the quasiparticle interaction vary 
strongly with the density up to $\rho \simeq \rho_0$. In particular the parameter
$f_0$ has a nearly linear depedence on the density, and at $k_f=1.7$\,fm$^{-1}$ gives
rise to a compressibility ${\cal K} = 560$\,MeV of neutron matter that grows strongly
with increasing density. The Landau parameter $f_1$ decreases rapidly at small densities, but 
flattens out close to nuclear matter saturation density, where it nearly vanishes thereby
yielding an effective mass $M^*/M_n \simeq 1$ according to eq.\ (\ref{effmass}). The $L=2$
component of the spin-independent quasiparticle interaction exhibits a nearly linear
decrease with the neutron density, owing to coherent effects from second-order two-nucleon
forces and three-neutron forces at leading-order.
The isotropic component of the spin-spin quasiparticle interaction, $g_0$, decreases with 
the neutron density, but we find no evidence for a phase transition to a ferromagnetic state
($G_0=N_0 g_0 \le -1$)
close to nuclear matter saturation density. For densities larger than $\rho_0$, the density-dependence of 
$g_0$ is governed mainly by the three-neutron force contributions, which decrease 
$g_0$ and consequently increase the neutron matter spin susceptibility $\chi$. 
The isotropic component $h_0$ of the exchange
tensor interaction is nearly independent of the density, since the medium corrections from two-
and three-body forces approximately cancel. The Fermi liquid parameters of the 
novel center-of-mass tensor and cross-vector interactions are relatively small in 
magnitude, but $k_0$, $l_0$ and $l_2$ have a strong density dependence.
From Table \ref{flpkf} we see that there remains a moderate dependence on the 
resolution scale and choice of low-energy constants. These variations lead to 
differences on the order of 10-30\% for most of the $L=0,1,2$ Fermi liquid parameters.

\begin{table}
\begin{tabular}{|c||D{.}{.}{2.3}|D{.}{.}{2.3}|D{.}{.}{2.3}||D{.}{.}{2.3}|D{.}{.}{2.3}|D{.}{.}{2.3}|}\hline
\multicolumn{1}{|c||}{$k_f=1.7$\,fm$^{-1}$} & \multicolumn{3}{c||}{Chiral N$^3$LO }
& \multicolumn{3}{c|}{$V_{\rm low-k}^{2.1}$} \\ \hline
\multicolumn{1}{|c||}{$L$} & \multicolumn{1}{c|}{0} & \multicolumn{1}{c|}{1} & 
\multicolumn{1}{c||}{2} & \multicolumn{1}{c|}{0} & 
\multicolumn{1}{c|}{1} & \multicolumn{1}{c|}{2} \\ \hline 
$f$ [fm$^2$] & 0.679 & -0.018 & -0.959 
& 1.072 & 0.135 & -0.686 \\ \hline 
$g$ [fm$^2$] & 1.025 & 0.388 & 0.302 
& 0.880 & 0.298 & 0.380  \\ \hline 
$h$ [fm$^2$] & 0.160 & 0.316 & -0.252 
& 0.175 & 0.317 & -0.263 \\ \hline
$k$ [fm$^2$] & -0.156 & 0.085 & 0.063 
& -0.108 & 0.051 & 0.039 \\ \hline
$l$ [fm$^2$] & -0.050 & -0.161 & -0.160 
& -0.295 & -0.330 & -0.029 \\ \hline
\end{tabular}
\caption{Fermi liquid parameters for the quasiparticle interaction in
neutron matter at a density corresponding to a Fermi momentum of $k_f=1.7$\,fm$^{-1}$. 
The low-energy constants of the N$^2$LO chiral three-nucleon force are chosen to
be $c_1 =-0.81\,$GeV$^{-1}$ and $c_3=-3.2\,$GeV$^{-1}$ when employed together with 
the bare chiral NN interaction and $c_1 =-0.76\,$GeV$^{-1}$ and $c_3=-4.78\,$GeV$^{-1}$
with the low-momentum interaction \vlkn.}
\label{flpkf}
\end{table}

%%%%%%%%%%%%%%%%%%%%%%%%%%%%%%%%%%%%%%%%%%%%%%%%%%%%%%%%%%%%%%%%%%%%%%%%%%%%%%%

%%%%%%%%%%%%%%%%%%%%%%%%%%%%%%%%%%%%%%%%%%%%%%%%%%%%%%%%%%%%%%%%%%%%%%%%%%%%%%%

\section{Conclusions and outlook}
In the present work we have computed the $L=0,1,2$ Fermi liquid parameters for 
the quasiparticle interaction in neutron matter employing realistic two-
and three-nucleon interactions derived within chiral effective field theory.
In addition to the free-space contribution from the two-body interaction, we have
calculated without any simplifying approximations the second-order two-body
corrrection as well as the leading three-body correction. 
A general method for extracting all components
of the quasiparticle interaction, both central and noncentral parts, 
for two-body interactions given in a partial-wave representation has been 
developed and tested with simple model interactions that can be treated semi-analytically.
Employing realistic two- and three-neutron forces, we find that medium-dependent 
loop corrections play an important role in increasing the compressibility of 
neutron matter from an unphysically small value to about ${\cal K}=600$\,MeV
at nuclear matter saturation density $\rho_0$. Second-order effects from 
two-body forces strongly enhance the quasiparticle effective mass $M^*$, while 
three-neutron forces play only a minor role for this quantity. 
The first- and second-order contributions to $g_0$ from two-body forces are
positive for the densities considered in the present work, though they decrease
in magnitude for increasing density. When combined with the medium-dependent loop
corrections from the leading-order chiral three-neutron forces, the Landau parameter $g_0$ decreases with
density, which leads to an increasing spin susceptibility $\chi$ of neutron matter.
The noncentral components of the quasiparticle interaction $h_L, k_L$ and $l_L$ have been computed 
as a function of the neutron density $\rho$. The extent to which they affect the 
spin susceptibility of neutron matter as well as the response functions
for electroweak probes will be studied in future work.

%%%%%%%%%%%%%%%%%%%%%%%%%%%%%%%%%%%%%%%%%%%%%%%%%%%%%%%%%%%%%%%%%%%%%%%%%%%%%%%

\section{Appendix: Energy per particle from model interactions at second order}

Organizing the second-order calculation of the energy density $\rho \bar E$
in the number of medium insertions, see 
eq.\ (\ref{imp}), one must evaluate (two-ring) Hartree and (one-ring) Fock diagrams each with 
two or three
medium insertions. We give first the pertinent analytical expressions for the pseudoscalar 
interaction in eq.\ (\ref{mopem}) at second order.

\noindent Hartree diagram with two medium insertions:
\begin{equation} \bar E(k_f)^{(H2)} = {g^4 M_n \over (8\pi)^3}
\bigg\{{21 \over 2u}
-15u +32 \arctan 2u -{7 \over 8
u^3}(3+20u^2)\ln(1+4u^2)\bigg\}\,,\end{equation}
Fock diagram with two medium insertions:
\begin{eqnarray} \bar E(k_f)^{(F2)} &=& {g^4 M_n \over (4\pi u)^3}
\bigg\{2u^2
-7u^3 \arctan u +6\sqrt{2}u^3 \arctan(\sqrt{2}u) \nonumber \\ &&
+\Big(1+{9u^2\over 2}\Big)\ln(1+u^2) -3 \Big({1\over 2}
+2u^2\Big)\ln(1+2u^2)
\nonumber \\ && + 4 \int_0^u\!\!dx \, x(u-x)^2(2u+x){1+4x^2+8x^4 \over
(1+2x^2)^3}
(\arctan x - \arctan 2x) \bigg\}\,,\end{eqnarray}
Hartree diagram with three medium insertions:
\begin{equation} \bar E(k_f)^{(H3)} = {g^4 M_n \over 32\pi^4
u^3}\int_0^u\!\!dx
\, x^2\!\int_{-1}^1\!\!dy \,\bigg[ 2ux y + (u^2-x^2y^2)\ln{u+x y \over u
- x y}
\bigg]{s^6 \over (1+s^2)^3}\,,
\end{equation}
Fock diagram with three medium insertions:
\begin{eqnarray} \bar E(k_f)^{(F3)} &=& {3g^4 M_n \over 64\pi^4
u^3}\int_0^u\!\!dx
\, \Bigg\{- 2\bigg[u - {1+u^2+x^2 \over 4x} \ln{1+(u+x)^2\over
1+(u-x)^2}\bigg]^2
+ x^2\!\int_{-1}^1\!\!dy\! \int_{-1}^1\!\!dz\nonumber \\ && \times { y z\,
\theta(y^2+z^2-1) \over |y z|\sqrt{y^2+z^2-1} }\bigg[\ln(1+s^2) - {s^2
\over 1+s^2}
\bigg] \bigg[\ln(1+t^2) - {t^2 \over 1+t^2}\bigg] \Bigg\}\,,\end{eqnarray}
with abbreviations $u=k_f/m$, $s=x y +\sqrt{u^2-x^2+x^2y^2}$ and $t=x z +\sqrt{u^2-x^2+x^2z^2}$.

For the spin-orbit interaction, eq.\ (\ref{soi}), at second order the analogous
expressions read

\begin{equation} \bar E(k_f)^{(H2)} = {g_s^4 M_n \over (4\pi)^3}
\bigg\{{5u \over 2}-
{2u^3 \over 3}-{3 \over 4u}- 4 \arctan 2u +{3+28u^2 \over 16 u^3}\ln(1+4u^2)
\bigg\}\,,
\end{equation}
\begin{eqnarray} \bar E(k_f)^{(F2)} &=& {g_s^4 M_n \over (2\pi u)^3}
\bigg\{{u^2
\over 2}-4u^3 \arctan u +3\sqrt{2}u^3 \arctan(\sqrt{2}u) \nonumber \\ &&
+(1+3u^2)\ln(1+u^2) -3 \Big({1\over 4} +u^2\Big)\ln(1+2u^2) \nonumber \\ &&
+ 2 \int_0^u\!\!dx \, x(u-x)^2(2u+x){1+4x^2+8x^4 \over
(1+2x^2)^3}(\arctan x -
\arctan 2x) \bigg\}\,,
\end{eqnarray}
\begin{eqnarray} \bar E(k_f)^{(H3)} &=& {g_s^4 M_n \over 64\pi^4
u^3}\int_0^u\!\!dx
\, x^2\!\int_{-1}^1\!\!dy \,\bigg[ 2ux y \Big(
{5u^2 \over 3} +2x^2-3x^2y^2 \Big) \nonumber \\ && + (u^2-x^2y^2)
(u^2+2x^2-3x^2y^2)
\ln{u+x y \over u - x y} \bigg]{s^4(3+s^2) \over
(1+s^2)^3}\,,\end{eqnarray}
\begin{eqnarray} \bar E(k_f)^{(F3)} &=& {3g_s^4 M_n \over 16\pi^4
u^3}\int_0^u\!\!dx
\, \Bigg\{- \bigg[u - {1+u^2+x^2 \over 4x} \ln{1+(u+x)^2\over
1+(u-x)^2}\bigg]^2
+ x^2\!\int_{-1}^1\!\!dy\! \int_{-1}^1\!\!dz\nonumber \\ && \times { y z\,
\theta(y^2+z^2-1) \over |y z|\sqrt{y^2+z^2-1} }\bigg[\ln(1+s^2) - {s^2
\over 1+s^2}
\bigg] \bigg[\ln(1+t^2) - {t^2 \over 1+t^2}\bigg] \Bigg\}\,,\end{eqnarray}
where $u=k_f/m_s$.
An interesting feature of the second-order spin-orbit interaction is
that for large mass $m_s$ the Hartree and Fock contributions become equal with
the same sign. It is worth noting that the agreement between these semi-analytic
results and those based on the partial wave decomposition eq.\ (\ref{ea2n}) 
agree on the per mille level.

%%%%%%%%%%%%%%%%%%%%%%%%%%%%%%%%%%%%%%%%%%%%%%%%%%%%%%%%%%%%%%%%%%%%%%%%%%%%%%%

\end{document}